%% file: paper_main.tex
\documentclass{IEEEtran}
\onecolumn
\usepackage{multirow}
\usepackage{subfigure}
\usepackage{graphicx}
\usepackage{longtable}
\usepackage{courier}
\usepackage{float}
\usepackage{amsmath,amssymb}

\usepackage[in]{fullpage}
\usepackage{draftwatermark}
\SetWatermarkLightness{0.9}

\newcommand{\Complex}{\mathbf{C}}

\title{Distortion bounds and Two-Way Protocols 
\\ for One-Shot
Transmission of \\
Correlated Random Variables}

\begin{document}
\author{Ay\c{s}e \"{U}nsal, Raymond Knopp\\
Mobile Communications Department, Eurecom, Sophia Antipolis, France\\
{\small{\texttt{ayse.unsal@eurecom.fr,
raymond.knopp@eurecom.fr}}}}
\maketitle
\thispagestyle{plain}
%% PLACE YOUR ABSTRACT HERE
\begin{abstract}
  \input{abstract}
\end{abstract}

\pagestyle{plain}
%% AND NOW START WITH YOUR PAPER CONTENT
\input{1_introduction}

\input{2_model}
\input{3_outer_bounds_sum}

\input{4_outer_bounds_parallel}
\input{5_single_source}
\input{6_dual_source}
\input{8_numerical}
\input{9_conclusion}

\input{10_Appendix}

%%%%%%%%%%%%%%%%%%%%%%%%%%%%%%%%%%%%%%%%%%%%%%%%%%%%%%%%%%%%%%%%%%
\bibliography{Corr_bib}
\bibliographystyle{IEEEtran}

%%%%%%%%%%%%%%%%%%%%%%%%%%%%%%%%%%%%%%%%%%%%%%%%%%%%%%%%%%%%%%%%%%

\end{document}

%% file: abstract.tex
This paper provides lower bounds on the reconstruction error for transmission of two continuous correlated random vectors sent over both sum and parallel channels using the help of two causal feedback links from the decoder to the encoders connected to each sensor. This construction is considered for both uniformly and normally distributed sources with zero mean and unit variance. Additionally, a two-way retransmission protocol, which is a non-coherent adaptation of the original work by Yamamoto is introduced for an additive white Gaussian noise channel with one degree of freedom. Furthermore, the novel protocol of a single source is extended to the dual-source case again for two different source distributions. Asymptotic optimality of the protocols are analyzed and upper bounds on the distortion level are derived for two-rounds considering two extreme cases of high and low correlation among the sources. It is shown by both the upper and lower-bounds that collaboration can be achieved through energy accumulation. Analytical results are supported by numerical analysis for both the single and dual-source cases to show the improvement in terms of distortion to be gained by retransmission subject to the average energy used by protocol . To cover a more realistic scenario, the same protocol of a single source is adapted to a wireless channel and their performances are compared through numerical evaluation.

\begin{keywords}
Distributed communication, joint source channel coding, correlation, multiple acces channel (MAC)
\end{keywords}

%% file: 1_introduction.tex
\section{Introduction \label{sec:intro}}
\footnote{This paper was presented [in part] at EUSIPCO 2012, European Signal Processing
Conference, August, 27-31, 2012, Bucharest, and SCC 2013, 9th International ITG Conference on Systems, Communications and Coding, January 21-24, 2013, Munich, Germany.}
In this work we consider simple transmission strategies for a network of sensors
able to measure a physical phenomenon from
different locations. Furthermore, we envisage a scenario where sensors operate under
tight energy constraints over a wireless transmission medium which motivates the use
of low-latency coding method. The
key issue is that digital transmission for small amounts of
typically analog data will induce overhead which is wasteful,
especially for massive networks of simple nodes. 
%Joint source-channel coding (JSCC), which combines the efforts of the channel and source code, 
%addresses such problems. In this paper, we consider JSCC for transmission of multiple spatially distributed
%samples of a slowly time-varying random field.

To illustrate this more precisely, imagine the simplest scenario of one sensor node tracking a slowly time-varying random sequence and sending its observations to a receiver over a wireless channel. The source is denoted by a random variable $U$ of zero mean and variance $\sigma_u^2=1$, representing a single realization of the random sequence at a particular time $t$. The sensor should be seen as a tiny device with strict energy constraints. The communication channel between the sender and the receiver is an additive white Gaussian noise channel. An important question is how to efficiently encode the random variable $U$ for transmission, 
and what performance can be achieved upon reconstruction as a function of the energy used to achieve this transmission. As an example, the sensor could be sporadically sending analog information
(temperature, magnetic field, current, speed, etc.) to a collecting node.  The traffic would be very 
low-rate (vanishing) and potentially requiring low-latency.  The latter could arise for two reasons, either reactivity 
of an actuating element in the network
or to minimize energy consumption in the sensing node itself by using discontinuous transmission and reception.  Here
the latency of the transmission is directly related to the ``on''-time of communication circuitry of the sensing node.
This example captures the essense of some so-called {\em machine-type communications}, a term which refers to 
machines (including sensors) interconnected via cellular networks and exchanging information autonomously.
  
For this scenario, the slowly time-varying characteristic of the source has two main impacts on the way the coding problem should be addressed: firstly, the time between two observations is long, and the sensor should not wait for a sequence of observations to encode it. Therefore, the sensor will encode only one observation before sending it through the channel. Secondly, for each source realisation the channel can potentially be used over many signal dimensions, for instance by encoding over a wide-bandwidth in the frequency-domain.  This would be the case for sensors connected directly to fourth-generation cellular networks.  Hence, we can reasonably assume that there is no constraint on the dimensionality of the channel codebook. The latter condition amounts to saying that very low-rate codes should be used.

\begin{figure}
\centering
\includegraphics[width=0.50\linewidth]{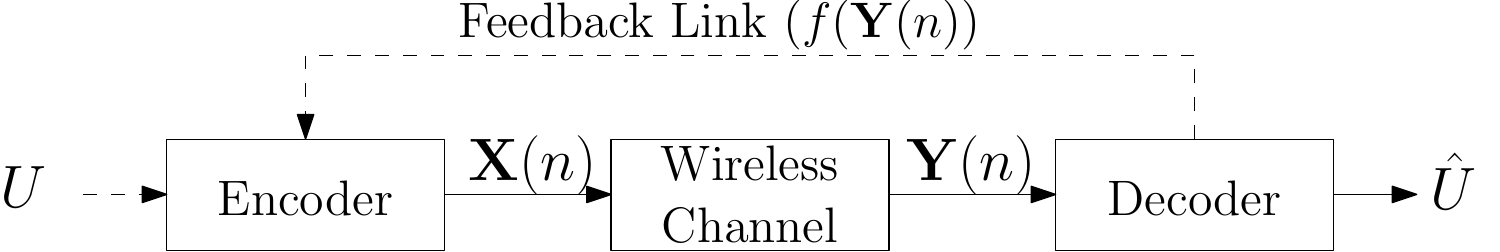}
\caption{Single-Source System Model}
\label{fig:system1}
\end{figure}

%We focus our attention on the case
%where unitary samples of the source are transmitted sporadically
%due to slow time-variation, and consequently we cannot
%perform sequence coding. 
The single-source model is depicted in Fig. \ref{fig:system1} where an encoder maps one realization of the source $U$ into $\mathbf{X} \triangleq (X_1,\ldots,X_N)$ where $N$ denotes the dimension of the channel input. We will make use of causal feedback so that the encoder may also depend on past channel outputs, that is $X_i=f(U,Y_1,\cdots,Y_{i-1})$. $\mathbf{X}$ is then sent across the channel corrupted by a white Gaussian noise sequence $\mathbf{Z}$, and is received as $\mathbf{Y}$. The receiver is a mapping function which tries to construct an estimate $\widehat{U}$ of $U$ given $\mathbf{Y}$. The fidelity criterion that we wish to minimize is the MSE distortion defined as $D\triangleq \mathbb{E}[(U-\widehat{U})^2]$,
under the mean energy constraint $\mathbb{E}[||\mathbf{X}||^2]\leq {\mathcal{E}}$. It is well-known that the linear encoder (i.e. $X=\sqrt{\mathcal{E}}U$) achieves the best performance under the mean energy constraint for the special case $N=1$ \cite{goblick}, \cite{elias}, \cite{gasp-thesis}. %An important generalization the case of multiple sensing nodes with spatially-correlated information as shown in Fig. \ref{fig:multisensor}. 
In fact, a lower bound on the distortion over all possible encoders and decoders is easily derived  in \cite{goblick} using classical information theory, and given by 
\begin{equation}
D\geq e^{-2\mathcal{E}/N_0} \label{eq:jscc-lb}
\end{equation}
where $N_0/2$ is the variance of the channel noise per dimension. \cite{schalkwijk67} achieves the same exponential behaviour through an achievable scheme for a band-limited Gaussian source in the presence of a noiseless feedback link.

An example of such a feedback-scheme for transmitting small amounts of information would be the random-access procedure in fourth-generation cellular networks, where a 6-bit message is conveyed using a orthogonal signal set occupying a large physical bandwidth.  The so-called {\em random-access response} contains the message hypothesized by the decoder, among other information, which serves either as an acknowledgement or an indication to retransmit.  Although simplified, we propose a scheme along these lines for the transmission of analog samples.
It is also inspired by Yamamoto's protocol \cite{Yamamoto79} which is an adaptation of the Schalkwijk-Barron scheme \cite{Schalkwijk71}.
%\begin{figure}[htbp]
  %\begin{minipage}[b]{0.5\linewidth}
   % \centering
   % \includegraphics[width=\linewidth]{Figures/system1}
   % \caption{System Model}
   % \label{fig:system1}
  %\end{minipage}
  %\hspace{0.5cm}
  %\begin{minipage}[b]{0.5\linewidth}
    %\centering
    %\includegraphics[width=\linewidth]{Figures/multisensor}
   % \caption{Multi-sensor Sampling and Transmission of a Random-Field}
  %  \label{fig:multisensor}
 % \end{minipage}
%\end{figure}

\begin{figure}
\centering
\includegraphics[width=0.50\linewidth]{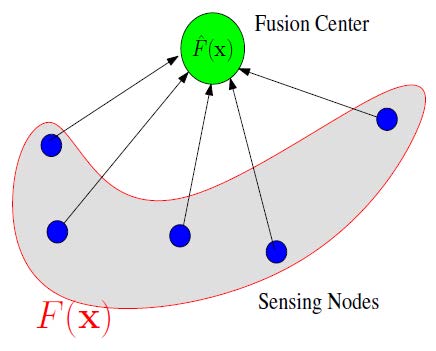}
\caption{Multi-sensor Sampling and Transmission of a Random-Field}
\label{Fig:multisensor}
\end{figure}

The multi-sensor scenario reflected in Figure \ref{Fig:multisensor} is an important generalization which is also considered here.  In particular we are interested in the case where two correlated random variables are transmitted over multiple-access channels using a similar scheme to the one described in Figure \ref{fig:system1}.  The key element being to exploit the correlation, which is assumed to be known, both at the transmitter and receiver.  Moreover, we aim to determine the operating regimes for such a multiple-access system in terms of the role correlation plays in determining the energy efficiency. In a similar vein, the authors in \cite{lapidoth1} and \cite{lapidoth2} derive a threshold signal-to-noise ratio (SNR) through the correlation between the sources so that below this threshold, minimum distortion is attained by uncoded transmission in a Gaussian multiple access channel with and without feedback, respectively. In these works, the authors consider transmission of a bi-variate normal source and the distortion can be characterized by two regimes as a function of the relationship between the channel SNR and the source SNR. Through a different approach lower bounds for transmission of correlated sources over Gaussian multiple-access channels is considered in \cite{Rajiv}.

It is important to note that in our scenario we are driven to assume unknown channels (i.e. non-coherent reception) in the formulation of the problem. Since the information content is very small, additional overhead for channel estimation is not warranted and thus, it is unreasonable to assume the channel state (i.e. channel amplitude and phase) be known to either the transmitter or receiver.  In what follows, simplifying steps in derivation of lower-bounds will result in equivalent formulations for known channels, however the proposed schemes will not make use of channel state information at either end of the transmission chain. 

\subsection{Outline and Contributions}
In section \ref{sec:model} we describe the source and channel models for the addressed problem. We consider two different models which induce correlation between the source vectors characterized by both uniform and normal statistics. Furthermore, two multiple-access channel models are used, namely a sum-channel and parallel-channel. We then provide lower-bounds on the reconstruction error for estimating two correlated continuous random-vectors transmitted across asynchronous multiple-access channels with feedback under different source and channel configurations in sections \ref{sec:sumch},\ref{sec:loosepar}.  The asymptotic behaviour of the obtained bounds is analyzed with respect to the level of correlation between the sources. In particular, we show that there are two regimes of operation characterizing the reconstruction error as a function of the energy used by the sensors.  One regime allows the collaboration through accumulation of energy from both sources, while the other does not.  In section \ref{sec:Novel-feedback}, we introduce a feedback scheme combining scalar quantization and orthogonal modulation which is applied to both the single and dual-sensor cases, for both uniform and Gaussian source variables.  We provide upper-bounds to the reconstruction error for this scheme and show that two regimes of operation can also be expected, although the relationship between correlation and energy used across the channel is different from what is predicted by the derivation of the lower-bounds.  In section \ref{sec:numerical} we finally provide numerical evaluations of both the lower and upper-bounds in order to draw conclusions on the efficiency of the proposed feedback scheme in comparison to the lower-bounds and to the case where a single transmission is used without feedback. We consider both non-coherent AWGN channels for both the upper and lower bounds and non-coherent fading channels for the upper-bounds. We show that collaboration can be achieved for a high-correlation regime with the proposed scheme, but that the gap between the lower-bound can be significant in the multi-sensor case.  In all cases, the benefit from feedback is very significant compared to a similar transmission scheme without exploiting feedback.

%% file: 2_model.tex
\section{Model Descriptions \label{sec:model}}
\subsection{Channel models}
Let us begin with the definition of the system models used to analyse the addressed problem together with the source-distribution and channel types. 
%Source vectors $\mathbf{U_{1},U_{2}}$ have a dimension of $K$ identically independent distributed samples of the correlated sources $U_{1},U_{2}$. 
\begin{figure}[htp]
  \centering
  \includegraphics[width=0.60\linewidth]{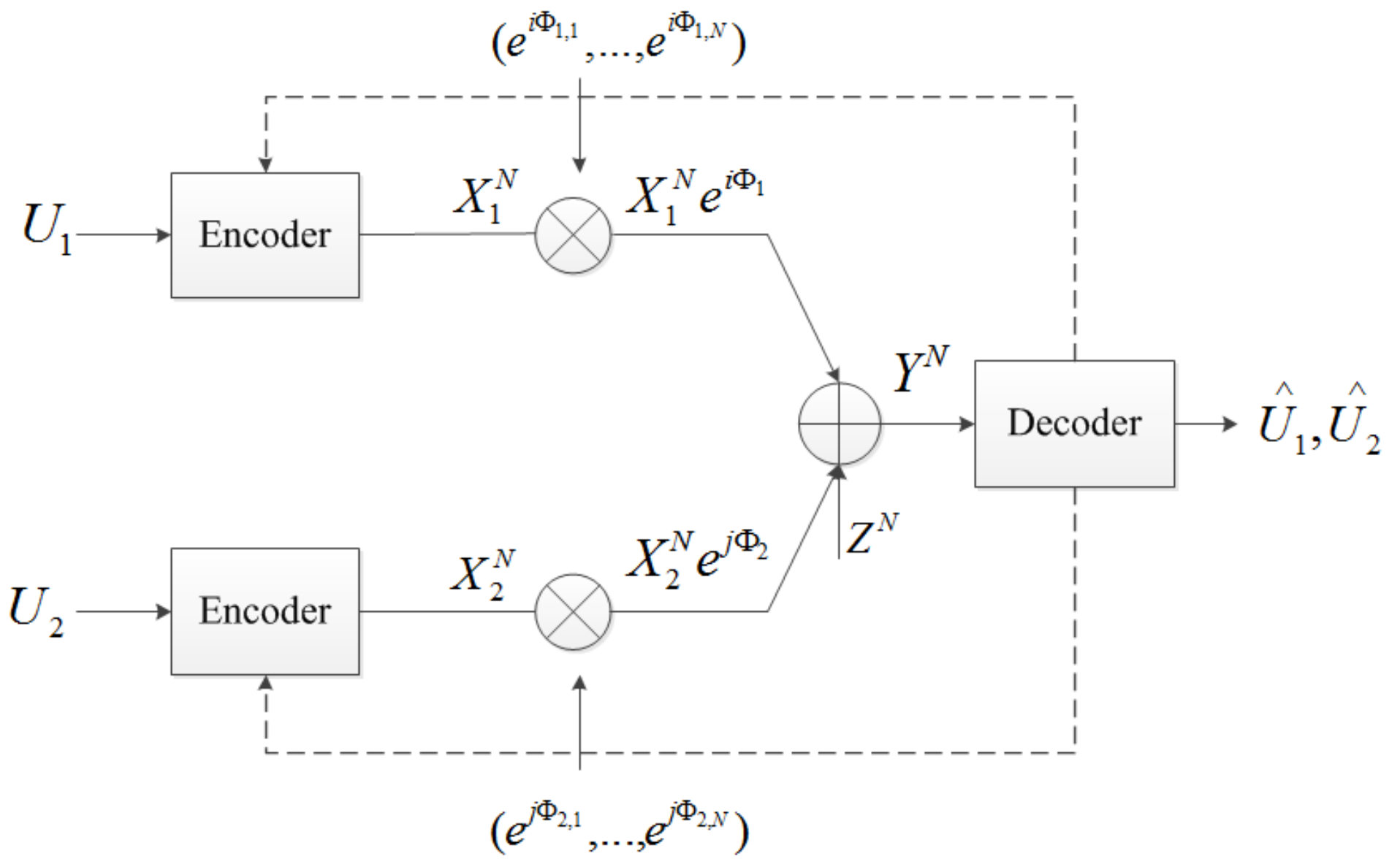}
  \caption{Correlated sources over GMAC with feedback.}
\label{fig:sources}
\end{figure}

The considered system for the sum-channel case is depicted in Figure \ref{fig:sources}.
The received signal $\mathbf{Y}=\{Y_{i};i=1,...,N\}$ and the power constraints are given as 
\begin{equation} \label{eq:dual-out}
Y_{i}=X_{1,i}e^{i\phi_{1,i}}+X_{2,i}e^{j\phi_{2,i}}+Z_{1,i}+Z_{2,i}
\end{equation}  
\begin{equation}
\frac{1}{K}\sum_{i=1}^{N} E[|X_{m,i}|^2]\leq \mathcal{E}_{m}
\end{equation}
for $m=1,2$ and $i,j=1,...,N$, respectively. The criteria to satisfy is chosen as the squared-error distortion measure, which is $d(u_{m},\hat{u}_{m})=(u_{m}-\hat{u}_{m})^2$. $\mathbf{\phi}_{m}=\{\phi_{m,i};i=1,...,N\}$ denotes the random phases which are assumed to be unknown both to the transmitter and the receiver.

\begin{figure}[htp]
  \centering
  \includegraphics[width=0.60\linewidth]{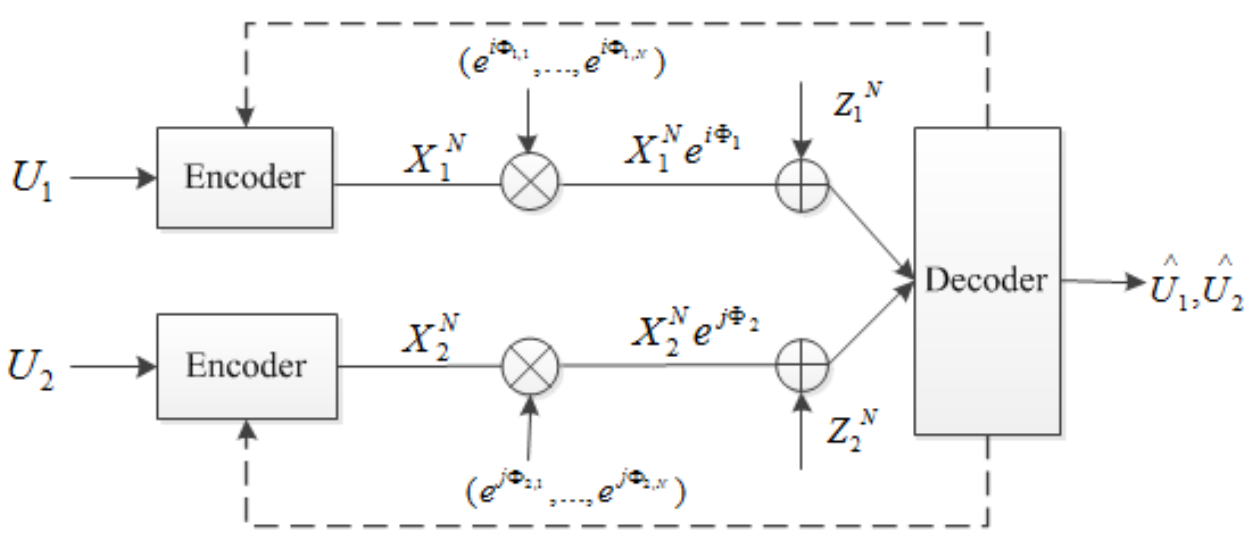}
  \caption{Transmission over parallel channels with feedback.}
\label{fig:parch}
\end{figure}

The second channel model under consideration is the parallel channel which is depicted in Figure \ref{fig:parch}
with the output signals and power constraints given below
\begin{align} \label{outputs-par}
Y_{1,i}&=X_{1,i}e^{i\phi_{1,i}}+Z_{1,i} \nonumber \\
Y_{2,i}&=X_{2,i}e^{j\phi_{2,i}}+Z_{2,i}
\end{align}
\begin{equation}
\frac{1}{K}\sum_{i=1}^{N} E[|X_{m,i}|^2]\leq \mathcal{E}_{m}
\end{equation}
for $m=1,2$ and $i=1,2,...,N$, respectively. ${\Phi}_{m}=\{\Phi_{m,i};i=1,...,N\}$ denotes the random phases which are assumed to be unknown both to the transmitter and the receiver as in the sum-channel model and the channel noise is defined as $\mathbf{Z}_{m}\sim N(0,N_{0}\mathbf{I}_{N})$. 

\subsection{Source models}

In the first case, which will be referred as source model I, the correlational relationship between the sources $\mathbf{U}_{1},\mathbf{U}_{2}$ dimension of $K$ is defined through the following expression
\begin{equation} \label{eq:sourcers}
\mathbf{U}_{2}=\rho \mathbf{U}_{1}+\sqrt{1-\rho^2}\mathbf{U}_{2}'
\end{equation}
where we denote the first source with $\mathbf{U}_{1}$ and the second source with $\mathbf{U}_{2}$. $\mathbf{U}_{2}'$ here is an auxiliary random vector. For the distributions of the two sources, two different types will be considered. In the first case, $\mathbf{U}_{1}$ is defined to be uniformly distributed over $(-\sqrt{3},\sqrt{3})$ and the second source $\mathbf{U}_{2}$ is defined to have a contaminated uniform distribution from(\ref{eq:sourcers}), based on $\mathbf{U}_{1}$ and $\mathbf{U}_{2}'$ which is also uniform on $(-\sqrt{3},\sqrt{3})$. So, we have one uniform and one near-uniform source having covariance equal to the correlation coefficient $\rho$ between them.  Secondly, in order to cover a more general case, the correlated sources $\mathbf{U}_{1}$ and $\mathbf{U}_{2}$ are defined to be standard normal random vectors, guaranteed by the auxiliary random vector $\mathbf{U}_{2}'$ is also normally distributed with zero mean and unit variance. Note that above definition is independent of the channel type. 

For the second source model, source model II, the two sources are defined by
\begin{equation} \label{eq:sources1}
\mathbf{U}_{1}=\rho \mathbf{U}+\sqrt{1-\rho^2}\mathbf{U}_{1}'
\end{equation}
\begin{equation}\label{eq:sources2}
\mathbf{U}_{2}=\rho \mathbf{U}+\sqrt{1-\rho^2}\mathbf{U}_{2}'
\end{equation}
where $\mathbf{U}$ is a mutual random vector and $\mathbf{U}_{1}'$ and $\mathbf{U}_{2}'$ are auxiliary random vectors dimension of $K$, which could be either uniformly distributed within the range $(-\sqrt{3},\sqrt{3})$ or normally distributed with zero mean and unit variance. Unlike source model I, this time both sources become contaminated uniform based on the level of the correlation. 
Although bounds on the distortion level of each source can be found for both models, results are given here only for source model I. We aim to observe the effect of the channel type both for the sum channels and the parallel channels covering two extreme cases of high and low correlation levels. The bounds for source model II can be found in \cite{LOLA_D4.4}, which are identical with the results of the first model in parallel channels whereas show slight difference only in terms of the factors in front in sum channels but not in the exponential behaviour.  Furthermore, source model II was be used in a subsequent work on large networks where the aim is to estimate the common random variable $U$ \cite{EURECOM+4068}.

\subsection{Discussion}

 In order to highlight the essense of the behaviour of the general case, we consider first the special case of a single source $\mathbf{U}$ dimension of $K$, whose message is sent over a Gaussian channel by being split into two parts through two different codebooks. Let us call the encoded parts of $\mathbf{U}$ as $\mathbf{X}_{1}$ and $\mathbf{X}_{2}$. The estimate $\mathbf{\hat{U}}$ is received after $\mathbf{X}_{1}$ and $\mathbf{X}_{2}$ are merged again before being decoded. In the following, $I(\mathbf{U};\mathbf{\hat{U}})$ is derived using two different expansions and the corresponding distortion $D$ is lower bounded. 
\begin{equation} \label{mutinf10'}
I(\mathbf{U};\mathbf{\hat{U}})\leq N\log \left(1+\frac{K\mathcal{E}}{NN_0}\right)
\end{equation}
%\begin{equation} \label{mutinf9'}
%I(\mathbf{U};\mathbf{\hat{U}})\geq \frac{K}{2}\log(\frac{1}{D})
%\end{equation}
and also
\begin{equation} \label{mutinf9'}
I(\mathbf{U};\mathbf{\hat{U}})\geq h(\mathbf{U})-h(\mathbf{U}-\mathbf{\hat{U}})
\end{equation} which varies based on the source distribution, since the entropy is directly related to the distribution type. The derivations of (\ref{mutinf10'}) and (\ref{mutinf9'}) are provided in Appendix \ref{sec:discussion}.
Combining above given two expansions, we get the lower-bound on distortion as 
\begin{equation} \label{ineq:asympd3}
D\geq C_{d}(1+\frac{K\mathcal{E}}{NN_0})^{-\frac{2N}{K}}
\end{equation}
which predicts that the energy used by the two transmitters can be accumulated. 
Asymptotically, letting $N\to \infty$ in (\ref{ineq:asympd3}) yields $D\geq C_d e^{-\frac{2\mathcal{E}}{N_0}}$ where $C_d$ is a constant defined by
\begin{equation} \label{constantd}
C_{d}=
\begin{cases}
\frac{6}{\pi e} & \text{Uniform source,}\\
1 & \text{Gaussian source.}\\
\end{cases}
\end{equation}
 In the upcoming sections \ref{sec:sumch}, \ref{sec:loosepar} and \ref{sec:tightpar}, it is shown that benefiting from the correlation between the sources, it is possible to achieve the behaviour of (\ref{ineq:asympd3}) and also a similar the energy efficiency with two highly correlated sources. 

%% file: 3_outer_bounds_sum.tex
\section{Distortion Bounds in a sum channel \label{sec:sumch}}

After describing the models for channel types, source constructions and distributions we introduce outer bounds on reconstruction error considering two extreme levels of correlation. This section covers the sum channel and provides lower bounds on distortion levels of each source. In order to avoid the repetations as giving the derivations of the outer bounds, we will use the notation $m$ to represent one of the sources and $m'$ will be used to indicate the other source, explicitly $m$ and $m'$ cannot be equal to 1 or 2 at the same time, when $m$ equals 1 then $m'$ has to be equal to 2 or vice versa.

Throughout the paper, the two different expansions of a mutual information one of which is based on the output signal and the other one is based on the sources are derived and equated in order to obtain a lower bound on the distortion level. For that reason, the first expansions on the output signals are applicable for both source distributions. Naturally, the second expansions vary depending on the source distribution.

\subsection{High correlation \label{sec:highcorrsumch}} 

%\subsubsection{Source model I \label{subsec:sum_high_I}}
In this part, we derive a relatively simple mutual information between the $m^{th}$ source $U_{m}$ and the output signal $Y$ through two different expansions considering the case where the sources are highly correlated, i.e. the correlation coefficient $\rho$ has a value close to 1.

In order to obtain a lower bound on the distortion level, two different expansions of $I(\mathbf{U}_{m};\mathbf{Y})$ are used considering the extreme case of highly correlated sources. First expansion of the desired mutual information based on the output signal is given by
\begin{equation}
I(\mathbf{U}_{m};\mathbf{Y})\leq N\log \left(1+\frac{K(\mathcal{E}_{m}+\mathcal{E}_{m'})}{N N_0}\right).\label{ineq:A'}
\end{equation} 
Same mutual information was derived through a different expansion and given by
\begin{equation}
I(\mathbf{U}_{m};\mathbf{Y})\geq h(\mathbf{U}_{m})-h(\mathbf{U}_{m}-\mathbf{\hat{U}}_{m})
\end{equation}
The derivations of both expansions given above can be found in Appendix \ref{sec:app_sum_high} together with the source entropies for $m=1$ and $m=2$.
Equating the two expansions of the same mutual information provides the below given bound on distortion level for the $m^{th}$ source
\begin{equation}\label{ineq:gendist}
D_{high,m}\geq C_{high,m} \left(1+\frac{K(\mathcal{E}_{m}+\mathcal{E}_{m'})}{NN_{0}}\right)^{-\frac{2N}{K}}
\end{equation}
where $C_{high,m}$ is a constant defined as
\begin{equation} \label{constant}
C_{high,m}=
\begin{cases}
\frac{6}{\pi e} & \text{for}\;\;\text{Uniform}\\
%\frac{6(1+\rho \sqrt{1-\rho^2})}{\pi e} & \text{if}\;\;m=2.\\
1 & \text{for}\;\;\text{Gaussian}\\
\end{cases}
\end{equation} %for uniform/contaminated uniform case whereas for the case where both sources are normally distributed, $C_{high,m}=1$.
Asymptotically in $N$, (\ref{ineq:gendist}) is obtained as
\begin{equation}\label{ineq:asym_hc_m}
D_{high,m}\geq C_{high,m} e^{-\frac{2(\mathcal{E}_{m}+\mathcal{E}_{m'})}{N_{0}}}.
\end{equation}

Note that, all bounds given above ( $C_{high,m}$ for $m=1,2$ and two different source distribution) have the same asymptotic behaviour independently of the source distributions and bring out the correlation benefit by using the sum energy of the two sources. 

Additionally, the product distortion term $D_{p}$ is bounded as given in the following form
\begin{equation}
D_{p}\geq C_{p}\exp\left(-\frac{2(\mathcal{E}_{m}+\mathcal{E}_{m'})}{N_{0}}\right), \label{ineq:product}
\end{equation}
with
\begin{equation} \label{constantIII}
C_{p}=
\begin{cases}
\frac{36\left(1-\rho^2\right)}{e^2 \pi^2} & \text{Uniform}\\
\left(1-\rho^2\right) & \text{Gaussian}\\
\end{cases}
\end{equation}
The derivation of the above given bound (\ref{ineq:product}) can be found in Appendix \ref{subsec:productb}. Next, we will observe the change on this behaviour based on the decrease in the correlation coefficient for the same channel type.
%\subsubsection{Source model II \label{subsec:sum_high_II}}

%The lower bound on the reconstruction error for estimating $U_{m}$ for a high level of correlation is obtained as given in the following
%\begin{equation}\label{ineq:gendistII}
%D_{m}\geq C_{m} \left(1+\frac{K(\mathcal{E}_{m}+\mathcal{E}_{m'})}{NN_{0}}\right)^{-\frac{2N}{K}}
%\end{equation}
%where $C_{m}$ is a constant defined as
%\begin{equation} \label{constantII}
%C_{m}=
%\begin{cases}
%1 & \text{Gaussian}\\
%\frac{6+12\rho \sqrt{1-\rho^2}}{\pi e} & \text{Uniform}\\
%\end{cases}
%\end{equation}
%The asymptotic of (\ref{ineq:gendistII}) is obtained as
%\begin{equation}\label{ineq:asym_hc_mII}
%D_{m}\geq C_{m} e^{-\frac{2(\mathcal{E}_{m}+\mathcal{E}_{m'})}{N_{0}}}.
%\end{equation}
%As noted before, the output signal expansion (\ref{ineq:A'}) of $I(\mathbf{U}_{m};\mathbf{Y})$ is applicable also to this model for both distribution types. On the contrary, for a different construction of the sources, second expansion of $I(\mathbf{U}_{m};\mathbf{Y})$ on the sources alters as shown in Appendix \ref{sec:app_sum_highII}. Note also that, source model II allows us to provide same bound for each source unlike the first model, in other words it is symmetric in this sense.

\subsection{Low correlation \label{sec:lowcorrsumch}}  
%\subsubsection{Source model I \label{subsec:sum_low_I}}
The main difference between this case and the previous one treated high correlation is the mutual information term to be used to bound the distortion level corresponding each source. Hence the mutual information between the source $\mathbf{U}_{m}$ and the output signal $\mathbf{Y}$ will be expanded through two different ways when the other source $\mathbf{U}_{m'}$ and the both phases are given. The two expansions of $I(\mathbf{U}_{m};\mathbf{Y}|\mathbf{U}_{m'},\Phi_{m},\Phi_{m'})$ are given as 
\begin{equation} \label{ineq:F'}
I(\mathbf{U}_{m};\mathbf{Y}|\mathbf{U}_{m'},\Phi_{m},\Phi_{m'})\leq N\log \left(1+\frac{K \mathcal{E}_{m}}{N N_{0}}\right),
\end{equation}
\begin{equation}
I(\mathbf{U}_{m};\mathbf{Y}|\mathbf{U}_{m'},\Phi_{m},\Phi_{m'})\geq h(\mathbf{U}_{m}|\mathbf{U}_{m'})-h(\mathbf{U}_{m}-\hat{\mathbf{U}}_{m}). \label{ineq:G'}
\end{equation} The derivations of (\ref{ineq:F'}) and (\ref{ineq:G'}) are given in Appendix \ref{sec:app_sum_low}.
The general form of the distortion bound  is obtained as $D_{low,m}\geq C_{low,m}\left(1+\frac{KE_{m}}{NN_{0}}\right)$ and asymptotically it becomes
\begin{equation} \label{ineq:asym_uni_lc_m}
D_{low,m}\geq C_{low,m}e^{-\frac{2\mathcal{E}_{m}}{N_{0}}},
\end{equation} where $C_{low,m}$ is a constant varying based on the source distribution and given by
\begin{equation} \label{constant_low}
C_{low,m}=
\begin{cases}
\frac{36(1-\rho^2)}{\pi^2 e^2} & \text{if}\;\;m=1,\\
\frac{6(1-\rho^2)}{\pi e} & \text{if}\;\;\; m=2\\
\end{cases}
\end{equation}
The normal distribution allows us to provide a single bound for both of the sources as
\begin{equation}
D_{low,m}\geq (1-\rho^2)e^{-\frac{2\mathcal{E}_m}{N_0}}.\label{ineq:low-sc-gauss-m}
\end{equation}

%\subsubsection{Source model II \label{subsec:sum_low_II}}
%The lower bound on the reconstruction error in estimating $U_{m}$ for a low correlation level is obtained as given in the following
%\begin{equation} \label{ineq:uni_lc_mII}
%D_{m}\geq C_{m}\left(1+\frac{K\mathcal{E}_{m}}{NN_{0}}\right)^{\frac{-2N}{K}},
%\end{equation} where $C_{m}$ is given by
%\begin{equation} \label{constant_lowII}
%C_{m}=
%\begin{cases}
%(1-\rho^4) & \text{Gaussian},\\
%\frac{6}{\pi e}(1-\rho^2)(1+\rho \sqrt{6/{\pi e}})^2 & \text{Uniform},\\
%\end{cases}
%\end{equation}which yields the asymptotic
%\begin{equation} \label{ineq:asym_uni_lc_mII}
%D_{m}\geq C_{m}e^{-\frac{2\mathcal{E}_{m}}{N_{0}}},
%\end{equation}
%It is clear also for low levels of correlation, second source model is symmetric in the sense that it provides one single bound for each distribution type. The derivation of (\ref{ineq:asym_uni_lc_mII}) can be found in Appendix \ref{sec:app_sum_lowII}.

For the Gaussian case, the lower bound can be given in the following general form for the $m^{th}$ source ensured by the symmetry of the problem
\begin{equation} \label{ineq:maxd2_g}
D_{m}\geq
\begin{cases}
D_{high,m} & \text{if}\;\;1-\rho^2 \leq \min(D_{m'},e^{-\frac{2\mathcal{E}_{m'}}{N_0}}),\\
D_{low,m} & \text{if}\;\;D_{m'}\geq e^{-\frac{2\mathcal{E}_{m'}}{N_0}} \; \mathrm{and} \; 1-\rho^2 \geq e^{-\frac{2\mathcal{E}_{m'}}{N_0}},\\
D_{p}/D_{m'}& \text{if}\;\;1-\rho^2 \geq \min(D_{m'},e^{-\frac{2\mathcal{E}_{m'}}{N_0}}).\\
\end{cases}
\end{equation}
On the other hand for the uniform case, the lower bounds on distortion $D_1$ and $D_2$ for the first and second source are given respectively as
\begin{equation} \label{ineq:maxd2_u1}
D_{1}\geq
\begin{cases}
D_{high,1} & \text{if}\;\;\frac{6(1-\rho^2)}{\pi e} \leq \min(D_{2},e^{-\frac{2\mathcal{E}_2}{N_0}}),\\
D_{low,1} & \text{if}\;\;D_{2}\geq e^{-\frac{2\mathcal{E}_2}{N_0}} \; \mathrm{and} \; \frac{6(1-\rho^2)}{\pi e} \geq e^{-\frac{2\mathcal{E}_2}{N_0}},\\
D_{p}/D_{2}& \text{if}\;\;\frac{6(1-\rho^2)}{\pi e} \geq \min(D_{2},e^{-\frac{2\mathcal{E}_2}{N_0}}),\\
\end{cases}
\end{equation}
\begin{equation} \label{ineq:maxd2_u2}
D_{2}\geq
\begin{cases}
%D_{high,2} & \text{if}\;\;\frac{1-\rho^2}{1+\rho \sqrt{1-\rho^2}} \leq \min((\pi e D_{1})/6,e^{-\frac{2\mathcal{E}_1}{N_0}}),\\
D_{high,2} & \text{if}\;\; 1-\rho^2 \leq \min((\pi e D_{1})/6,e^{-\frac{2\mathcal{E}_1}{N_0}}),\\
D_{low,2} & \text{if}\;\; D_{1}\geq \frac{6}{\pi e }e^{-\frac{2\mathcal{E}_1}{N_0}} \; \mathrm{and} \; 1-\rho^2 \geq e^{-\frac{2\mathcal{E}_1}{N_0}},\\
D_{p}/D_{1}& \text{if}\;\; 1-\rho^2 \geq \min((\pi e D_{1})/6,e^{-\frac{2\mathcal{E}_1}{N_0}}).\\
\end{cases}
\end{equation}
%\begin{equation} \label{ineq:maxd2}
%D_{m}\geq \max(D_{m,low}(\mathcal{E}_m),D_{m,high}(\mathcal{E}_m+\mathcal{E}_m'),\sqrt{D_{m}D_{m'}(\mathcal{E}_m+\mathcal{E}_m')})
%\end{equation}
%where we denote (\ref{ineq:asym_uni_lc_m}) or (\ref{ineq:low-sc-gauss-m}) by $D_{m,low}(\mathcal{E}_m)$, (\ref{ineq:asym_hc_m}) by $D_{m,high}(\mathcal{E}_m+\mathcal{E}_m')$ and (\ref{ineq:product}) by $D_{m}D_{m'}(\mathcal{E}_m+\mathcal{E}_m')$. 
The above bounds predict that energy accumulation cannot be achieved when the distortion resulting from the estimation of 
one source realization using the other (i.e. $1-\rho^2$) is more than the point-to-point distortion (Goblick bound \cite{goblick}) incurred during transmission.

%% file: 4_outer_bounds_parallel.tex
\section{Distortion Bounds in parallel channels \label{sec:loosepar}} 

\subsection{High Correlation}
%\subsubsection{Source model I \label{subsec:Par_high_I}}
Let us consider the model described in Section \ref{sec:model} and depicted in Figure (\ref{fig:parch}). The aim is to bound the distortion level of each source as the sum-channel model studied in the previous section. To begin with, consider the use of two different expansions of $I(\mathbf{U}_{m};\mathbf{Y}_{m},\mathbf{Y}_{m'})$. The first expansion based on the channel output signals is given by
\begin{equation}
I(\mathbf{U}_{m};\mathbf{Y}_{m},\mathbf{Y}_{m'})\leq N\log \left(1+\frac{K\mathcal{E}_{m}}{NN_{0}}\right)\left(1+\frac{K\mathcal{E}_{m'}}{NN_{0}}\right). \label{ineq:L'}
\end{equation}
Note that the expression given above is independent of the source distribution unlike the second expansion given by
\begin{equation}
I(\mathbf{U}_{m};\mathbf{Y}_{m},\mathbf{Y}_{m'})\geq h(\mathbf{U}_{m})-h(\mathbf{U}_{m}-\mathbf{\hat{U}}_{m}). \label{ineq:M'}
\end{equation} 
Equating (\ref{ineq:L'}) and (\ref{ineq:M'}) provides the lower bound on distortion
\begin{equation}
D_{high,m}\geq C_{high,m}\left(1+\frac{K\mathcal{E}_{m}}{NN_{0}}\right)^{-\frac{2N}{K}}\left(1+\frac{K\mathcal{E}_{m'}}{NN_{0}}\right)^{-\frac{2N}{K}} \label{ineq:genbound}
\end{equation}
and its limiting expression $D_{high,m}\geq C_{high,m} e^{-\frac{2(\mathcal{E}_{m}+\mathcal{E}_{m'})}{N_{0}}}$ where $C_{high,m}$ is a constant given as
\begin{equation} \label{constant_p}
C_{high,m}=
\begin{cases}
\frac{6}{\pi e} & \text{for}\;\; \text{Uniform}\\
%\frac{6(1+\rho \sqrt{1-\rho^2})}{\pi e} & \text{if}\;\;m=2.\\
1 & \text{for}\;\; \text{Gaussian.}\\
\end{cases}
\end{equation}
Choosing the corresponding constant value from the function (\ref{constant_p}) provides the following distortion bounds are given as
\begin{equation}
D_{high,m}\geq C_{high,m}e^{-\frac{2(\mathcal{E}_{m}+\mathcal{E}_{m'})}{N_{0}}} \label{asym_2uni_hc_m}
\end{equation}
%\begin{equation}
%D_{1}\geq \frac{6}{\pi e} e^{-\frac{2(E_{1}+E_{2})}{N_{0}}}, \label{asym_2uni_hc}
%\end{equation}
%\begin{equation}
%D_{2}\geq \frac{6+12\rho \sqrt{1-\rho^2}}{\pi e}e^{-\frac{2(E_{1}+E_{2})}{N_{0}}}, \label{asym_2uni_hc2}
%\end{equation}
%whereas the factor $C_{high,m}$ equals to 1 for both sources given that the Gaussian entropies are equal to each other, hence we obtain a single bound for the two as
%\begin{equation}
%D_{high,m}\geq e^{-\frac{2(\mathcal{E}_m+\mathcal{E}_m')}{N_0}}\label{asym_2gauss_hc_m}.
%\end{equation}
The derivations of (\ref{ineq:L'}) and (\ref{ineq:M'}) can be found in Appendix \ref{sec:app_par_high}.

Additionally, the product distortion term for this channel construction yields the same asymptotic bound (\ref{ineq:product}) derived in the previous section of sum channel. The derivation can be found in Appendix \ref{sec:product_p}.
%\subsubsection{Source model II \label{subsec:Par_high_II}}

%Using the second source model, we do not gain only the same exponential behaviour for the extreme case of a high correlation in both channels, but also same factors $C_{m}$ as in hte previous channel type. So source model II in parallel channels provides (\ref{ineq:asym_hc_mII}) with the factor as given by (\ref{constantII}). Given the difference between the construction of the channels, the derivation of the mutual information $I(\mathbf{U}_{m};\mathbf{Y}_{m},\mathbf{Y}_{m'})$ based on the output signals differs slightly compared to the sum channel case whereas it is identical with (\ref{ineq:L'}), since the change of source model has no effect on it.
%On the other hand, the second expansion differs and is obtained as in the sum channel case. Thus the asymptotic bounds introduced for the second source model for a sum channel in case of a high correlation given by (\ref{ineq:asym_hc_mII}) apply to parallel channels as well.

\subsection{Low Correlation}
%\subsubsection{Source model I \label{subsec:Parallel_low_I}}
Let us evaluate another mutual information based on one source and its corresponding output signal given the other source and the corresponding output signal together with the random phases to observe the effect of correlation on the above derived bounds. In this case, $\rho$ is considered to be close to 0. The first expansion of the mutual information $I(\mathbf{U}_{m};\mathbf{Y}_{m}|\mathbf{U}_{m'},\mathbf{Y}_{m'})$ is as follows
\begin{equation}
I(\mathbf{U}_{m};\mathbf{Y}_{m}|\mathbf{U}_{m'},\mathbf{Y}_{m'})\leq N\log \left(1+\frac{K\mathcal{E}_{m}}{NN_{0}}\right) \label{ineq:O'}
\end{equation}
And the second expansion based on the sources is  given by in the general form 
\begin{equation} \label{ineq:N'}
I(\mathbf{U}_{m};\mathbf{Y}_{m}|\mathbf{U}_{m'},\mathbf{Y}_{m'})\geq h(\mathbf{U}_{m}|\mathbf{U}_{m'})-h(\mathbf{U}_{m}-\hat{\mathbf{U}}_{m})
\end{equation}
Using the two expansions (\ref{ineq:O}) and (\ref{ineq:N}) of $I(\mathbf{U}_{m};\mathbf{Y}_{m}|\mathbf{U}_{m'},\mathbf{Y}_{m'},\Phi_{m})$, the lower bound on the distortion in parallel channels with a low correlation between the two sources is obtained and given in the following general form of
\begin{equation}
D_{low,m}\geq C_{low,m}\left(1+\frac{K\mathcal{E}_{m}}{NN_{0}}\right)^{-\frac{2N}{K}}
\end{equation}
and the above bound becomes asymptotically in $N$, $D_{low,m}\geq C_{low,m}e^{-\frac{2E_m}{N_0}}$ where $C_{low,m}$ is defined as
\begin{equation} \label{constant2}
C_{low,m}=
\begin{cases}
%(1-\rho^2) & \text{for} \;\; \text{Gaussian},\\
\frac{36(1-\rho^2)}{\pi^2 e^2} & \text{if} \;\; m=1,\\
\frac{6(1-\rho^2)}{\pi e} & \text{for} \;\; m=2\\
\end{cases}
\end{equation}
and $(1-\rho^2)$ in Gaussian case for $m=1,2$. 
Using the relevant constants from (\ref{constant2}), the final form of the distortion bounds become
\begin{equation}
D_{low,m}\geq C_{low,m}e^{-\frac{2\mathcal{E}_{m}}{N_{0}}}. \label{asym_2uni_lc}
\end{equation}
On the other hand, we provide a single bound for normally distributed sources
\begin{equation}
D_{low,m}\geq (1-\rho^2)e^{-\frac{2\mathcal{E}_{m}}{N_{0}}}.\label{asym_2gauss_lc_m}
\end{equation}

%Based on the range of $1-\rho^2$, we have the following bounds for the uniform case in a parallel channel for the first and the second source, respectively.
%\begin{equation} \label{ineq:maxd2_u1_p}
%D_{1}\geq
%\begin{cases}
%D_{high,1} & \text{if}\;\;\frac{6(1-\rho^2)}{\pi e} \leq \min((\pi e D_{2})/6,e^{-\frac{2\mathcal{E}_2}{N_0}}),\\
%D_{low,1} & \text{if}\;\;D_{2}\geq e^{-\frac{2\mathcal{E}_2}{N_0}} \; \mathrm{and} \; \frac{6(1-\rho^2)}{\pi e} \geq e^{-\frac{2\mathcal{E}_2}{N_0}},\\
%D_{p}/D_{2}& \text{if}\;\;\frac{6(1-\rho^2)}{\pi e} \geq \min((\pi e D_{2})/6,e^{-\frac{2\mathcal{E}_2}{N_0}}),\\
%\end{cases}
%\end{equation}
%\begin{equation} \label{ineq:maxd2_u2_p}
%D_{2}\geq
%\begin{cases}
%D_{high,2} & \text{if}\;\;\frac{1-\rho^2}{1+2\rho \sqrt{1-\rho^2}} \leq \min( D_{1}/6,e^{-\frac{2\mathcal{E}_1}{N_0}}),\\
%D_{low,2} & \text{if}\;\; D_{1}\geq \frac{6}{\pi e }e^{-\frac{2\mathcal{E}_1}{N_0}} \; \mathrm{and} \; \frac{1-\rho^2}{1+2 \rho \sqrt{1-\rho^2}}  \geq e^{-\frac{2\mathcal{E}_1}{N_0}},\\
%D_{p}/D_{1}& \text{if}\;\;\frac{1-\rho^2}{1+2 \rho \sqrt{1-\rho^2}} \geq \min( D_{1}/6,e^{-\frac{2\mathcal{E}_1}{N_0}}).\\
%\end{cases}
%\end{equation}
Parallel channel yields the same result based on the range of $1-\rho^2$, as the sum channel for Gaussian case given by (\ref{ineq:maxd2_g}) whereas (\ref{ineq:maxd2_u1}) and (\ref{ineq:maxd2_u2}) are valid for the first and the second source in the uniform case, respectively.

%\subsubsection{Source model II  \label{subsec:Parallel_low_II}}

%As shown in the high correlation case, source model II yields the same exponential behaviour as the first model together with the correlational dependence indicated by the factor in front $C_m$. Thus lower bounds on the distortion level introduced in Subsection \ref{subsec:sum_low_II}, the expression (\ref{ineq:asym_uni_lc_mII}) applies to the current model. In other words, although the derivation differs slightly on the output side, source model II do not result differently among channel types. 

%Let us denote (\ref{asym_2uni_lc}) or for the gaussian case (\ref{asym_2gauss_lc_m}) by $D_{m,low}(\mathcal{E}_m)$ and (\ref{asym_2uni_hc_m}) or (\ref{asym_2gauss_hc_m}) by $D_{m,high}(\mathcal{E}_m,\mathcal{E}_m')$ so that the final bound on $D_{m}$ is given as a maximum function of these two; 
%\begin{equation}
%D_{m}\geq \max(D_{m,low}(\mathcal{E}_m),D_{m,high}(\mathcal{E}_m,\mathcal{E}_m'))
%\end{equation}
%Using this method the effect of the correlation on the distortion can only be evaluated through the maximum point of the two, one of which (\ref{asym_2uni_lc} or \ref{asym_2gauss_lc_m}) covers the case of low correlation and uses the individual energy of the corresponding source as the other bound (\ref{asym_2uni_hc_m} or \ref{asym_2gauss_hc_m}) holds for highly correlated sources which benefits from the sum energy of the sources. 

In the next subsection, we will give the derivation of a special bound only on $D_{2}$ for uniform case and both on $D_{1}$ and $D_{2}$ for the Gaussian case, which cover both of high and low correlation behaviour in a single expression which is why tighter than relatively to the former two.

\subsection{A relatively tighter alternative for parallel channels\label{sec:tightpar}} 

It will be shown that, using another expression for the mutual information to be used to derive the distortion bound in parallel channels, a relatively tighter bound compared to those given in the former section can be achieved only for the second source for uniform/contaminated uniform construction and for both sources when they are normally distributed. We start with the single bound for the uniform case and through deriving the mutual information $I(\mathbf{U}_{2};\mathbf{Y}_{2}|\mathbf{Y}_{1})$ we get the first expansion as
\begin{equation} \label{ineq:mutinf12'}
I(\mathbf{U}_{2};\mathbf{Y}_{2}|\mathbf{Y}_{1})\leq N\log(1+\frac{K\mathcal{E}_{2}}{NN_{0}})
\end{equation}
The second expansion of the same mutual information is given by
\begin{equation}
I(\mathbf{U}_{2};\mathbf{Y}_{2}|\mathbf{Y}_{1})\geq \frac{K}{2}\log \left(2^{\frac{2}{K}h(\sqrt{1-\rho^2}\mathbf{U}_{2}')}+2^{\frac{2}{K}(K\log|\rho|+h(\mathbf{U}_{1}|\mathbf{Y}_{1})}\right)-h(\mathbf{U}_{2}-\mathbf{\hat{U}}_{2}) \label{mutinf11'}
\end{equation}
where in step (a), we used the entropy-power inequality in order to expand the entropy $h(\rho \mathbf{U}_{1}+\sqrt{1-\rho^2}\mathbf{U}_{2}'|\mathbf{Y}_{1})$. Consequently, we obtain $h(\mathbf{U}_{1}|\mathbf{Y}_{1})$ in a general form as follows
\begin{equation} \label{parso'}
h(\mathbf{U}_{1}|\mathbf{Y}_{1})\geq h(\mathbf{U}_{1})-N\log \left(1+\frac{K\mathcal{E}_{1}}{NN_{0}}\right).
\end{equation}
We obtain the bound on $D_{2}$ 
\begin{equation} \label{pardist1}
D_{2}\geq \frac{6(1-\rho^2)}{\pi e}\left(1+\frac{K\mathcal{E}_{2}}{NN_{0}}\right)^{-\frac{2N}{K}}
+\frac{6\rho^2}{\pi e}\left[\left(1+\frac{K\mathcal{E}_{1}}{N N_{0}}\right)\left(1+\frac{K\mathcal{E}_{2}}{NN_{0}}\right)\right]^{-\frac{2N}{K}} 
%\\
%+\frac{12\rho \sqrt{1-\rho^2}}{\pi e}\left(1+\frac{K\mathcal{E}_{1}}{N N_0}\right)^{-\frac{N}{K}} \left(1+\frac{K\mathcal{E}_{2}}{NN_0}\right)^{-\frac{2N}{K}}
\end{equation}
and let $N\to \infty$ the bound given above becomes
\begin{equation}
D_{2}\geq \frac{6(1-\rho^2)}{\pi e}\exp \left(-\frac{2\mathcal{E}_{2}}{N_{0}}\right)+\frac{6\rho^2}{\pi e}\exp \left(-\frac{2(\mathcal{E}_{1}+\mathcal{E}_{2})}{N_{0}}\right)
%+\frac{6\rho \sqrt{1-\rho^2}}{\pi e}\exp \left(-\frac{\mathcal{E}_{1}+2\mathcal{E}_{2}}{N_{0}}\right).
\end{equation}
Secondly, for the Gaussian case, distortion level $D_{2}$ is bounded by
\begin{equation}
D_{2}\geq (1-\rho^2) \left(1+\frac{K\mathcal{E}_{2}}{NN_0}\right)^{-\frac{2N}{K}}+\rho^2 \left[\left(1+\frac{K\mathcal{E}_{1}}{NN_0}\right)\left(1+\frac{K\mathcal{E}_{2}}{NN_0}\right)\right]^{-\frac{2N}{K}}
\end{equation}
Let $N\to \infty$, above bound becomes
\begin{equation}
D_{2}\geq (1-\rho^2) e^{-\frac{2\mathcal{E}_{2}}{N_0}}+ \rho^2 e^{-\frac{2(\mathcal{E}_{1}+\mathcal{E}_{2})}{N_0}}.
\end{equation}
As noted above, distribution type allows us to achieve another expression for the first source through defining $\mathbf{U}_{1}=\frac{1}{\rho} \mathbf{U}_{1}+\frac{\sqrt{1-\rho^2}}{\rho}\mathbf{U}_{2}'$. Finally we obtain the following lower bound on $D_1$.
\begin{equation}
D_{1}\geq \frac{1-\rho^2}{\rho^2}\left(1+\frac{K\mathcal{E}_{2}}{NN_0}\right)^{-\frac{2N}{K}}+\frac{1}{\rho^2}\left[\left(1+\frac{K\mathcal{E}_{1}}{NN_0}\right)\left(1+\frac{K\mathcal{E}_{2}}{NN_0}\right)\right]^{-\frac{2N}{K}}
\end{equation}
Accordingly, asymptotic of the above bound is obtained as
\begin{equation}
D_{1}\geq \frac{1-\rho^2}{\rho^2} e^{-\frac{2\mathcal{E}_{1}}{N_0}}+ \frac{1}{\rho^2} e^{-\frac{2(\mathcal{E}_{1}+\mathcal{E}_{2})}{N_0}}.
\end{equation}
The derivations of all three bounds can be found in Appendix \ref{sec:app_par_D2}.

%% file: 5_single_source.tex
\section{Asymptotic Optimality of Simple Two-Way Protocol with Non-coherent Detection \label{sec:Novel-feedback}}

Let us consider now a non-coherent version of the Schalkwijk-Barron \cite{Schalkwijk71}/Yamamoto \cite{Yamamoto79} protocol applied to the transmission of isolated analog samples with non-coherent reception. 
This will serve as a motivating example for the use of feedback with low-latency achieving asymptotically near-optimal
distortion performance.  In the analysis, we first focus on a simple AWGN channel with a one dimensional source letter.  

The protocol consists of two phases, a data phase and a control phase. In our adaptation the two phases compose one round of the protocol. A source sample quantized to $B$ bits is encoded into one of $2^B$ $N$-dimensional messages $\mathbf{S}_{m}$, with $m=1,2,\cdots$ and each message is transmitted with equal energy $\sqrt{\mathcal{E}_{\mathrm{D},i}}$, where $\mathcal{E}_{\mathrm{D},i}$ denotes the energy of the data phase on the $i^{th}$ round. Upon reception, the receiver computes the maximum-likelihood 
(or MAP if source is non-uniform) message, $\hat{m}(\mathbf{Y}_d)$, based on the $N$-dimensional observation
\begin{equation}
\mathbf{Y}_d = \sqrt{\mathcal{E}_{\mathrm{D},i}}\mathrm{e}^{j\Phi_i}\mathbf{S}_m + \mathbf{Z} \label{eq:y_d}
\end{equation}
where the subscript $d$ represents the current phase. 
The random phase sequence $\phi_i$ is assumed to be i.i.d. with uniform distribution on $[0,2\pi)$. The $N$-dimensional
vector noise sequence $\mathrm{z}_i$ is complex, circularly symmetric, has zero-mean and 
autocorrelation $N_0\mathbf{I}_{N\times N}$. After the first data phase, the receiver 
feeds $\hat{m}$ back to the encoder via the noiseless feedback link.  Let the 
corresponding error event be denoted $E_i$.
After the data phase, the encoder enters the control phase and informs the receiver whether or not its decision 
was correct via a signal $\sqrt{\mathcal{E}_{\mathrm{C},i}}\mathbf{S}_c$ of energy $\sqrt{\mathcal{E}_{\mathrm{C},i}}$ if the
decision is incorrect and $\mathbf{0}$ if the decision was correct. $\mathcal{E}_{\mathrm{C},i}$ here denotes the energy of the control phase in the $i^{th}$ round. During the control phase the receiver observes
$\mathbf{Y}_{c}$. Let $y_c=\mathbf{Y_c}^H\mathbf{S}_c$ and assume a detector of the form
\begin{equation}
\mathrm{e}=\mathcal{I}\left(|y_c|^2 > \lambda\mathcal{E}_{\mathrm{C},i}\right) \label{eq:detector}
\end{equation}
where $\mathcal{I}(\cdot)$ is the indicator function and $\lambda$ is a threshold to be optimized and included within the interval $[0,1)$.
%Let the error events be denoted $E_{\mathrm{c}\rightarrow\mathrm{e},1}$ and $E_{\mathrm{e}\rightarrow\mathrm{c},1}$as described and analyzed in \cite{Yamamoto79}. 
As described and analyzed in \cite{Yamamoto79}, $E_{\mathrm{e}\rightarrow\mathrm{c},i}$ corresponds to an uncorrectable error since it acknowledges an error as correct decoding and $E_{\mathrm{c}\rightarrow\mathrm{e},i}$ represents a misdetected acknowledged error declaring correct decoding as incorrect. 
If the receiver correctly decodes the control signal and it signals that the data phase was correct after the completion of the first round, with
probability $\Pr(E_1^c)(1-\Pr(E_{\mathrm{c}\rightarrow\mathrm{e},1}))$, the protocol
halts, otherwise another identical round is initiated by the receiver.  The retransmission probability, i.e. the probability of going on for a second round, is $\Pr(E_1)(1-\Pr(E_{\mathrm{e}\rightarrow\mathrm{c},1}))$. This on-off signaling guarantees that with
probability $\Pr(E_1^c)(1-\Pr(E_{\mathrm{c}\rightarrow\mathrm{e},1}))$ the transmitter will not expend more than $\mathcal{E}_{\mathrm{D},1}$ joules, which should be close to one. After each data phase, the receiver computes the ML or MAP message
$\hat{m}_i(\mathbf{Y}_1,\cdots,\mathbf{Y}_i)$ based on all observations up to round $i$ with error event 
$E_i$.  
The same control phase is repeated and the protocol is terminated after $N$ rounds. 
The reconstruction error of the source message is obtained by calculating the mean squared error distortion through
\begin{equation}
D=D_{q}(1-P_{e})+D_{e}P_{e} \label{eq:distgen}
\end{equation}
and can be bounded further as
\begin{equation}
D\leq D_{q}+D_{e}P_{e} \label{ineq:distgen}
\end{equation}
where $P_{e}$ is the total probability of error, $D_{q}$ represents the distortion caused by the quantization process and $D_{e}$ corresponds to the MSE distortion for the case where an error was made. 
The error probability at the end of round $N$ is defined and consequently bounded by
\begin{align} \label{eq:peN_gen}
&P_{\mathrm{e}}=\sum_{i=1}^{N-1} \Pr(E_i)\Pr(E_{\mathrm{e}\rightarrow\mathrm{c},i}) \prod_{i=1}^{N-1}(1-\Pr(E_{\mathrm{e}\rightarrow\mathrm{c},i}))\nonumber \\
&+\sum_{j=0}^{2^{N-1}} \Pr(j) \prod_{i=0}^{N-1}(\Pr(E_{\mathrm{c}\rightarrow\mathrm{e},i}))^{1-B_i(j)}(1-\Pr(E_{\mathrm{e}\rightarrow\mathrm{c},i}))^{B_i(j)}Pr(E_N|j)\nonumber \\
&\overset{(a)}{\leq} \sum_{i=1}^{N-1} \Pr(E_i)\Pr(E_{\mathrm{e}\rightarrow\mathrm{c},i}) + \Pr(E_N)
\end{align}
where $B_{i}(j)=\mathcal{I}(\mathrm{Round}\; j\; \mathrm{in}\; \mathrm{error})$ and in step (a) the conclusive expression is obtained through bounding $\Pr(E_{\mathrm{e}\rightarrow\mathrm{c},i})$ and $(1-\Pr(E_{\mathrm{e}\rightarrow\mathrm{c},i}))$ by 1.
Average energy used by the protocol of $N$ rounds is 
\begin{equation} \label{eq:aven_N}
\mathcal{E}=\mathcal{E}_{\mathrm{D},1}+ \sum_{i=2}^{N}\mathcal{E}_{\mathrm{D},i} \left[\Pr(E_{i-1})(1-\Pr(E_{\mathrm{e}\rightarrow\mathrm{c},i-1}))\right] 
+\sum_{i=2}^{N}\mathcal{E}_{\mathrm{D},i} \left[(1-\Pr(E_{i-1}))\Pr(E_{\mathrm{c}\rightarrow\mathrm{e},i-1})\right] 
+ \sum_{i=1}^{N-1}\Pr(E_{i})\mathcal{E}_{\mathrm{C},i}
\end{equation}

The probability of an uncorrectable error in round $i$ is obtained as
\begin{align}
\Pr(E_{\mathrm{e}\rightarrow\mathrm{c},i}) &= \Pr\left(|\sqrt{\mathcal{E}_{\mathrm{C},i}} + z_c|^2 \leq \lambda\mathcal{E}_{\mathrm{C},i}\right)\nonumber\\
&=1-\mathrm{Q}_1\left(\sqrt{\frac{2\mathcal{E}_{\mathrm{C},i}}{N_0}},\sqrt{\frac{2\lambda\mathcal{E}_{\mathrm{C},i}}{N_0}}\right)\label{eq:MQfunc},
\end{align}
where $\mathrm{Q}_{1}(\alpha,\beta)$ is the first-order Marcum-$\mathrm{Q}$ function and
$z_c=\mathbf{S}_c^H\mathbf{Z}$ is a circularly-symmetric Gaussian zero-mean random variable 
with variance $N_0$.  Furthermore, we have the recent bound on the $\mathrm{Q}_{1}(\alpha,\beta)$ 
for $\alpha>\beta$ from \cite[eq:4]{Simon} which is very useful for bounding (\ref{eq:MQfunc}) as follows
\begin{equation}
\Pr(E_{\mathrm{e}\rightarrow\mathrm{c},i})\leq 1/2 \exp \left(-\frac{(\sqrt{\lambda}-1)^2 \mathcal{E}_{\mathrm{C},i}}{N_0}\right) 
%&\leq \frac{\mathrm{arc sin}\left(\sqrt{\lambda}\right)}{\pi}
%\left[\exp\left(-\left(1+\lambda-2\sqrt{\lambda}\right)\frac{\mathcal{E}_{\mathrm{C},1}}{N_0}\right)-\right.\nonumber\\
%& \;\;\;\;\;\;\;\;\left.\exp\left(-\left(1+\lambda+2\sqrt{\lambda}\right)\frac{\mathcal{E}_{\mathrm{C},1}}{N_0}\right)\right] \nonumber\\&\leq \frac{\mathrm{arc sin}\left(\sqrt{\lambda}\right)}{\pi}
%\exp\left(-\left(1+\lambda-2\sqrt{\lambda}\right)\frac{\mathcal{E}_{\mathrm{C},1}}{N_0}\right) 
\label{eq:pec}
\end{equation}
The probability of a misdetected acknowledged error is obtained as
\begin{align}
\Pr(E_{\mathrm{c}\rightarrow\mathrm{e},i}) &= \Pr\left(|z_c|^2 > \lambda\mathcal{E}_{\mathrm{C},i}\right)\nonumber\\
&= \mathrm{e}^{-\frac{\lambda\mathcal{E}_{\mathrm{C},i}}{N_0}} \label{eq:prob_c_e}
\end{align}
Lastly, the probability of making an error on a particular round $L$, $\Pr(E_L)\leq  2^{B}P_2(L)$ can be derived using \cite[eq:12.1-24]{ProakisBook}
\begin{equation} \label{eq:prgen}
P_2(L)\leq\frac{1}{2^{2L-1}}e^{-\gamma /2}\sum_{n=0}^{L-1}c_n\left(\frac{\gamma}{2}\right)^n
\end{equation}
where $c_n=1/n! \sum_{k=0}^{L-1-n}\dbinom{2L-1}{k}$ and $\gamma$ represents the signal to noise ratio.

\subsection{Performance of Two-rounds} \label{subsec:tworounds}
%In the following, we will consider the case where the protocol is repeated up to two rounds, i.e. $N=2$. 
In this part, the resulting the probability of error is investigated together with the average energy used by protocol and the reconstruction error considering that the protocol is repeated for two rounds, i.e. $N=2$. 
%The probability of error for the case where the protocol is repeated up to $N$ rounds is bounded by
The probability of error at the end of the second round is defined and bounded as 
\begin{align}
P_{\mathrm{e}}^{(2)} &= \Pr(E_1)\Pr(E_{\mathrm{e}\rightarrow\mathrm{c},1}) + \Pr(E_1)(1-\Pr(E_{\mathrm{e}\rightarrow\mathrm{c},1}))\Pr(E_2|E_1)+ (1-\Pr(E_1))\Pr(E_{\mathrm{c}\rightarrow\mathrm{e},1})\Pr(E_2|E_1^c)\nonumber\\
&\leq \Pr(E_1)\Pr(E_{\mathrm{e}\rightarrow\mathrm{c},1}) + \Pr(E_2) \label{eq:pe}
\end{align}
which is obtained through (\ref{eq:peN_gen}) with $N=2$. Here, $P_{\mathrm{e}}^{(2)}$ represents the total probability of error at the end of the second round (\ref{eq:peN_gen} with $N=2$)
The average energy used by the protocol is
\begin{equation}
\mathcal{E}= \mathcal{E}_{\mathrm{D},1} + \Pr(E_1)\mathcal{E}_{\mathrm{C},1} + (\Pr(E_1)(1-\Pr(E_{\mathrm{e}\rightarrow\mathrm{c},1}))+(1-\Pr(E_1))\Pr(E_{\mathrm{c}\rightarrow\mathrm{e},1}))\mathcal{E}_{\mathrm{D},2}. \label{eq:avener}
\end{equation}
$\mathcal{E}_{\mathrm{D},2}$ here denotes the required energy for retransmission, which is the energy to be used in the data phase of the second round.
Clearly if $\Pr(E_{\mathrm{e}\rightarrow\mathrm{c},1})$ and $\Pr(E_{\mathrm{c}\rightarrow\mathrm{e},1})$ are 
small, then the protocol achieves marginally more than $\mathcal{E}_{\mathrm{D},1}$ joules per source symbol. The detection rule is given using \cite[Chapter 12, eq:12.1-16]{ProakisBook} considering the following 2 possible decision variables assuming $(k)$ is transmitted.
\begin{equation}
U_{k}=|\sqrt{\mathcal{E}_{D,1}}+N_k|^2 \label{dv1_single}
\end{equation}
\begin{equation}
U_{k'}=|N_{k'}|^2 \label{dv2_single}
\end{equation}
where $U_{k}=|<\mathbf{Y}_1,\mathbf{S}_{m_{k}}>|^2$. An error is committed if  $U_{k'}$ is greater than $U_{k}$. The union bound on $P_{e}(k)$ is defined as 
\begin{equation} \label{ineq:unionb}
P_{e}(k)\leq \sum_{(k')\neq(k)}\Pr\left(u_{k}<u_{k'}|(k)\right)
\end{equation} The conditional probability of $U_{k}<U_{k'}$  given $(k)$ is transmitted becomes for the first round
\begin{equation} \label{eq:probdv1}
\Pr(U_{k}<U_{k'}|k)=\Pr(U_{k}<U_{k'})=\Pr(|\sqrt{\mathcal{E}_{D,1}}+N_k|^2<|N_{k'}|^2)
\end{equation} whereas for the second round, we have cumulatively the following probability
\begin{equation} \label{eq:probdv3}
\Pr(U_{k}<U_{k'}|k)=\Pr(U_{k}<U_{k'})
=\Pr(|\sqrt{\mathcal{E}_{D,1}}+N_{k,1}|^2+\sqrt{\mathcal{E}_{D,2}}+N_{k,2}|^2<|N_{k',1}|^2+|N_{k',2}|^2)
\end{equation}
Bounds on the error probabilities of both rounds are attained through (\ref{eq:prgen}) and given by
\begin{equation}
\Pr(E_1)\leq 2^{B-1}\mathrm{e}^{-\frac{\mathcal{E}_{\mathrm{D},1}}{2N_0}}, \label{eq:pe1}
\end{equation}
\begin{equation}
\Pr(E_2) \leq 2^{B-3}\left(1+3\frac{\mathcal{E}_{\mathrm{D},1}+\mathcal{E}_{\mathrm{D},2}}{N_0}\right)\mathrm{e}^{-\frac{\mathcal{E}_{\mathrm{D},1}+\mathcal{E}_{\mathrm{D},2}}{2N_0}}. \label{eq:pe2}
\end{equation} where (\ref{eq:pe1}) corresponds to (\ref{eq:probdv1}) which is equivalent to $P_2(1)$ and (\ref{eq:pe2}) is obtained through (\ref{eq:probdv3}) equivalently by $P_2(2)$.
%\subsection{Reconstruction Error} \label{sec:dist}

The mean squared-error distortion for a uniform source $U$ on $(-\sqrt{3},\sqrt{3})$ 
%or a gaussian source with parameters $(0,1)$
, i.e. a source with zero mean and unit variance, is obtained as
\begin{equation}
D\left(\mathcal{E},N_0,N,\lambda\right)=2^{-2B}(1-P_e) + 2P_{e}. \label{eq:dist}
\end{equation}

In order to bound the reconstruction error (\ref{eq:dist}) and to observe its asymptotic performance, (\ref{ineq:distgen}) is applied to (\ref{eq:dist}) and combined with (\ref{eq:pe}), (\ref{eq:pec}) for $i=1$, (\ref{eq:pe1}) and (\ref{eq:pe2}). The resulting distortion is bounded as
\begin{equation}
D\left(\mathcal{E},N_0,2,\lambda\right) \leq K_1\mathrm{e}^{-2B\ln 2} 
+ K_2\mathrm{e}^{(B-1)\ln 2 - \frac{\mathcal{E}_{\mathrm{D},1}}{2N_0} - \left(1+\lambda-2\sqrt{\lambda}\right)\frac{\mathcal{E}_{\mathrm{C},1}}{N_0}}
+K_3\mathrm{e}^{(B-2)\ln 2 - \frac{\mathcal{E}_{\mathrm{D},1}+\mathcal{E}_{\mathrm{D},2}}{2N_0}}
\label{eq:distbound}\end{equation}
where $K_1$ and $K_2$ are $O(1)$, while $K_3$ is $O(\mathcal{E}_{\mathrm{D},1}+
\mathcal{E}_{\mathrm{D},2})$. 
By equating coefficients in the three exponentials of (\ref{eq:distbound}) we have that $\mathcal{E}_{\mathrm{C},1}=\frac{\mathcal{E}_{\mathrm{D},2}}{2(1+\lambda-2\sqrt{\lambda})}$. In order for $\Pr(E_1)$ to be very close to zero so that $\mathcal{E}$ can be made arbitrarily close to $\mathcal{E}_{\mathrm{D},1}$, we define $\mathcal{E}_{D,2}=(2-\mu)\mathcal{E}_{D,1}$ where $\mu$ is an arbitrary constant satisfying $\mu\in(0,2)$. Finally, we obtain the bound on the distortion at the end of the second round as given by
\begin{equation}
D\left(\mathcal{E},N_0,2\right)\leq K_D\mathrm{e}^{-\frac{\mathcal{E}_{\mathrm{D},1}(1+\mu/3)}{N_0}} \label{finaldist}
\end{equation}
with $K_D\sim O(\mathcal{E}_{\mathrm{D},1})$. It is worth mentioning, the limiting expression in \cite[eq.15]{schalkwijk67} is achieved 
to within a factor of 1/2 in the energy in two rounds with non-coherent reception.

\subsection{Third round and after \label{sec:third_r}}

Assume that the protocol is not terminated after the second round, so it goes on one more round to do the retransmission. Hereafter, we will show that the asymptotic performance (\ref{finaldist}) achieved in two rounds cannot be improved unless the average energy used by protocol is increased. The probability of error given by (\ref{eq:peN_gen}) can be simply bounded as in (\ref{eq:pe}) for $N=3$
\begin{equation}
P_{\mathrm{e}}^{(3)}\leq \sum_{i=1}^{2}\Pr(E_i)\Pr(E_{\mathrm{e}\rightarrow\mathrm{c},i})+ \Pr(E_3) \label{eq:pe3}
\end{equation}
%\begin{equation}
%P_{\mathrm{e}}\leq \sum_{l=1}^{2}\Pr(E_l)\Pr(E_{\mathrm{e}\rightarrow\mathrm{c},l})+ \Pr(E_3) \label{eq:pe3}
%\end{equation}
%where $\Pr(E_l)$ remains the same for $l=1,2$ and the cumulative error probability $\Pr(E_3)$ is defined as
with
\begin{equation}
\Pr(E_3) \leq 2^{B-5}\mathrm{e}^{-\frac{\mathcal{E}_{\mathrm{D},1}+\mathcal{E}_{\mathrm{D},2}+\mathcal{E}_{\mathrm{D},3}}{2N_0}}
\left(16+6\frac{\mathcal{E}_{\mathrm{D},1}+\mathcal{E}_{\mathrm{D},2}+\mathcal{E}_{\mathrm{D},2}}{N_0}+1/2 \left(\frac{\mathcal{E}_{\mathrm{D},1}+\mathcal{E}_{\mathrm{D},2}+\mathcal{E}_{\mathrm{D},2}}{N_0}\right)^2\right) \label{eq:pe4}
\end{equation} which is equivalent to $P_2(3)$ representing the cumulative error probability at the end of the third round where $\mathcal{E}_{\mathrm{D},3}$ denotes the energy used in the corresponding round.
%The uncorrectable error in one round is independent of the uncorrectable error in another round, so that (\ref{eq:pec}) can be used also for the second round with $\mathcal{E}_{\mathrm{C},2}$. 
The distortion at the end of third round is bounded as
\begin{align}
D\left(\mathcal{E},N_0,3,\lambda\right) &\leq K_1\mathrm{e}^{-2B\ln 2} + K_2\mathrm{e}^{(B+1)\ln 2 - \frac{\mathcal{E}_{\mathrm{D},1}+2 \left(1-\sqrt{\lambda}\right)^2\mathcal{E}_{\mathrm{C},1}}{2N_0}}\nonumber \\
&+K_3\mathrm{e}^{(B-1)\ln 2 - \frac{\mathcal{E}_{\mathrm{D},1}+\mathcal{E}_{\mathrm{D},2}+2 \left(1-\sqrt{\lambda}\right)^2\mathcal{E}_{\mathrm{C},2}}{2N_0}}
+K_4\mathrm{e}^{(B-3)\ln 2 - \frac{\mathcal{E}_{\mathrm{D},1}+\mathcal{E}_{\mathrm{D},2}+\mathcal{E}_{\mathrm{D},3}}{2N_0}} \label{eq:distbound3}
\end{align}
where $K_4=O((\mathcal{E}_{\mathrm{D},1}+\mathcal{E}_{\mathrm{D},2})^2)$. By equating the coefficients in the four exponentials of (\ref{eq:distbound3}), we obtain the following relationships between the energies $\mathcal{E}_{\mathrm{D},2}=\mathcal{E}_{\mathrm{D},3}=\mathcal{E}_{\mathrm{C},2}2(1-\sqrt{\lambda})^2$ and $\mathcal{E}_{\mathrm{C},1}=2\mathcal{E}_{\mathrm{C},2}$.
The final form of the upper bound on the distortion level at the end of the third round becomes
\begin{equation}
D\left(\mathcal{E},N_0,3\right)\leq K_{D_{3}}\mathrm{e}^{-\frac{\mathcal{E}_{\mathrm{D},1}(1-2\mu_{2}/3)}{N_0}} \label{finaldist3}
\end{equation}
where we defined $\mathcal{E}_{\mathrm{D},2}=\mathcal{E}_{\mathrm{D},3}=(1-\mu_2)\mathcal{E}_{\mathrm{D},1}$ to assure the average energy used by protocol for three rounds to be arbitrarily close to the energy  only in the first round. $\mu_2$ is an arbitrary constant satisfying $\mu_{2}\in(0,1)$. This result proves that the asymptotic performance achieved in two rounds cannot be improved with more rounds. 
%If the exponential behaviour of (\ref{finaldist3}) achieves $\exp \{-\frac{2\mathcal{E}_{\mathrm{D},1}}{N_0}\}$ (i.e. twice better than the performance achieved in two rounds) by changing the relationship between the energies used in three rounds, this causes the average energy used by the protocol to exceed $\mathcal{E}_{\mathrm{D},1}$, the energy used in the data phase of the first round. 
Moreover, even though it is possible to obtain $\exp \{-\frac{2\mathcal{E}_{\mathrm{D},1}}{N_0}\}$ (i.e. twice better than the performance in (\ref{finaldist}) by changing the relationship between the energies used in the different rounds, this causes the average energy used by the protocol to exceed $\mathcal{E}_{\mathrm{D},1}$, the energy used in the data phase of the first round.

%% file: 6_dual_source.tex
\section{Extension of the Protocol for Two Correlated Sources\label{sec:dual_gen}}

\begin{figure}[htp]
  \centering
  \includegraphics[width=0.50\linewidth]{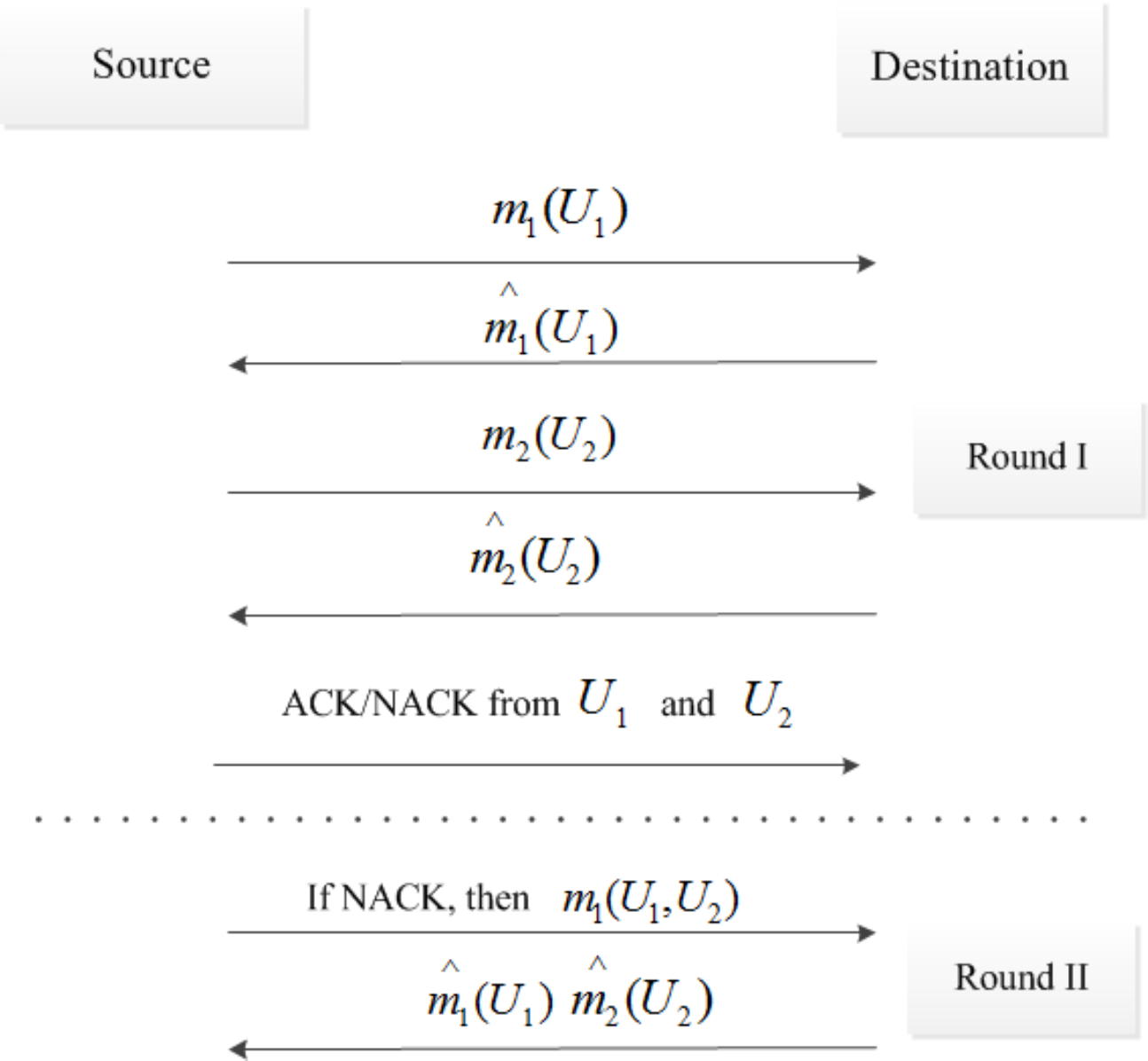}
  \caption{Two-round protocol}
\label{fig:protocol}
\end{figure}

The total energy to be used by protocol is fixed and we will denote the energy used in the data phase of the $i^{th}$ round by the $j^{th}$ source by $\mathcal{E}_{\mathrm{D},i,j}$, where $i,j=1,2$. In the same way, $\mathcal{E}_{\mathrm{C},i,j}$ denotes the energy used in the control phase of the $i^{th}$ round by the $j^{th}$ source. The quantized source sample of the $j^{th}$ source is encoded into $2^{B_{j}}$ messages with dimension $N$. 
The quantization processes are treated in detail in the following subsections separately since it differs based on the source distribution. 
In the data phase, the first source sends its message $\mathbf{m}_1(U_{1})$ to the receiver with energy $\mathcal{E}_{\mathrm{D},1,1}$. The receiver detects $\hat{m}_{1}$ and feeds it back. And the second source sends $\mathbf{m}_2(U_{2})$ with energy $\mathcal{E}_{\mathrm{D},1,2}$. The energy in the control phase of the $i^{th}$ round is defined as $\mathcal{E}_{\mathrm{C},i}= \mathcal{E}_{\mathrm{C},i,1}+\mathcal{E}_{\mathrm{C},i,2}$ and the total energy in the data phase is $\mathcal{E}_{\mathrm{D},i}= \mathcal{E}_{\mathrm{D},i,1}+\mathcal{E}_{\mathrm{D},i,2}$. This encoding rule allows the second source to exploit the correlation of its sample with that of its peer and the energy used is chosen according to the likelihood of the estimate fed back from the receiver. After the estimation and feedback of $\hat{m}_2$, data phase of the first round ends and the encoders enter the control phase to inform the receiver about the correctness of its decision, as in the single source case. For that, each source sends ACK/NACK signals regarding its own message to the decoder. According to the control signals, either the protocol halts or goes on another round to do the retransmission of the message which were not acknowledged in the control phase. For the second data phase, the destination instructs the sources to retransmit and re-detect its message. Proceeding of the protocol is illustrated in Figure \ref{fig:protocol}.
Extending the output signal based on the $N$ dimensional observation to the current scheme with dual-source, output signal of the $j^{th}$ source in the data phase is 
\begin{equation} \label{eq:output_s}
\mathbf{Y}_d = \sqrt{\mathcal{E}_{\mathrm{D},1,j}}\mathrm{e}^{j\Phi_j}\mathbf{S}_{m_{j}} + \mathbf{Z}_{j}.
\end{equation}
We assume the random phases $\Phi_{j}$ to be distributed uniformly on $[0,2\pi)$, the channel noise $\mathbf{Z}_{j}$ to have zero mean and equal autocorrelation $N_{0}\mathbf{I}_{N\times N}$ for $j=1,2$ and  $\mathbf{S}_{m_{j}}$ are the $N$-dimensional messages, where $m=1,2,\cdots,2^{B_j}$ and $j=1,2$. We have the same form of detector described in (\ref{eq:detector}) for the $j^{th}$ source as $e_{j}=I\left(|y_{c,j}|^{2}>\lambda_{j}\mathcal{E}_{\mathrm{C},1,j}\right)$ with $y_{c,j}=\mathbf{Y_{c,j}}^H\mathbf{S}_{c,j}$
$\lambda_{1}$ and $\lambda_{2}$ are threshold values to be optimized and included within the interval $[0,1)$. For simplification, we will assume $\lambda_{1}$ and $\lambda_{2}$ to be equal to the same value $\lambda$. 
We denote the error events in the first round and on the $j^{th}$source with $E_{1,j}$.

Let $e_{1,j}$ and $c_{1,j}$ denote erroneous and correct decoding in the first round on $U_{j}$, respectively. Accordingly $E_{c\rightarrow e,1}$ and $E_{e\rightarrow c,1}$ are used to denote a mis-detected acknowledged error and an uncorrectable error, respectively. 
The probability of an uncorrectable error in the first round is taken as the sum of the probability of errors of each source as
$\Pr(E_{e\rightarrow c,1})=\sum_{j=1}^{2}\Pr(E_{e\rightarrow c,1,j})$. 
The probability of an uncorrectable error $E_{e\rightarrow c}$ for $U_{j}$ is given by
\begin{align}
\Pr(E_{e \rightarrow c,1,j})&=\Pr(|\sqrt{\mathcal{E}_{\mathrm{C},1,j}}+z_{c,j}|^2 \leq \lambda\mathcal{E}_{\mathrm{C},1,j}) \nonumber \\
&=1-Q_{1}\left(\sqrt{\frac{\mathcal{E}_{\mathrm{C},1,j}}{N_{0}/2}},\sqrt{\frac{\lambda\mathcal{E}_{\mathrm{C},1,j}}{N_{0}/2}}\right)\nonumber \\
&\overset{(a)}{\leq}1/2 \exp\left(-\frac{(\sqrt{\lambda}-1)^2 \mathcal{E}_{\mathrm{C},1,j}}{N_{0}}\right). \label{probec}
\end{align}
using the recent bound on the $Q_{1}(\alpha,\beta)$ given in \cite[eq:4]{Simon} in step (a).
%\begin{equation}
%\Pr(E_{e \rightarrow c,1,j})\leq 1/2 \exp(-\frac{(\sqrt{\lambda}-1)^2 \mathcal{E}_{\mathrm{C},1,j}}{N_{0}}). \label{probec}
%\end{equation}
The total probability of a mis-detected acknowledged error to occur in the first round is obtained in the same way by;
$\Pr(E_{c\rightarrow e,1})=\sum_{j=1}^{2}\Pr(E_{c\rightarrow e,1,j})$.
And the probability of a mis-detected acknowledged error $E_{c\rightarrow e}$ for $U_{j}$ is 
\begin{equation}
\Pr(E_{c \rightarrow e,1,j})=\exp\left\{{-\frac{\lambda \mathcal {E}_{\mathrm{C},1,j}}{N_{0}}}\right\}. \label{misdetected}
\end{equation}
The protocol uses the average energy given by 
\begin{align} \label{eq:energy}
{\mathcal{E}}&=\mathcal{E}_{\mathrm{D},1,1}+\mathcal{E}_{\mathrm{D},1,2}+\mathcal{E}_{\mathrm{C},1,1}\Pr(E_{1,1},E_{1,2}^c)+\mathcal{E}_{\mathrm{C},1,2}\Pr(E_{1,1}^c,E_{1,2})+(\mathcal{E}_{\mathrm{C},1,1}+\mathcal{E}_{\mathrm{C},1,2})\Pr(E_{1,1},E_{1,2})\nonumber \\
&+\mathcal{E}_{\mathrm{D},2}[\Pr(E_{1,1},E_{1,2}^c)(1-\Pr(E_{e\rightarrow c,1,1}))+\Pr(E_{1,1}^c,E_{1,2})(1-\Pr(E_{e\rightarrow c,1,2}))\nonumber \\
&+\Pr(E_{1,1},E_{1,2})(1-\Pr(E_{e\rightarrow c,1,1})\Pr(E_{e\rightarrow c,1,2}))+\Pr(E_{1,1}^c,E_{1,2}^c)(\Pr(E_{c\rightarrow e,1,1})+\Pr(E_{c\rightarrow e,1,2}))]\nonumber \\
&\overset{(a)}{\leq} \mathcal{E}_{\mathrm{D},1}+ \frac{\mathcal{E}_{\mathrm{C},1}}{2} [\Pr(E_{1,1},E_{1,2}^c)+\Pr(E_{1,1}^c,E_{1,2})] +\mathcal{E}_{\mathrm{C},1}\Pr(E_{1,1},E_{1,2})\nonumber \\
&+\mathcal{E}_{\mathrm{D},2}[\Pr(E_{1,1},E_{1,2}^c)+\Pr(E_{1,1}^c,E_{1,2})+\Pr(E_{1,1},E_{1,2})+\Pr(E_{1,1}^c,E_{1,2}^c)\Pr(E_{c\rightarrow e,1})]
\end{align} In step (a), the probability of correct detection of an error $(1-\Pr(E_{e\rightarrow c,1,j}))$ and the probability of both sources being correct in the first round $\Pr(E_{1,1}^c,E_{1,2}^c)$ are upper bounded by 1, which suggests the average energy used by protocol can be made arbitrarily close to the energy used in the data phase of the first round for a vanishing error probability.  

The detection rule is given using \cite[Chapter 12]{ProakisBook} considering the following 4 possible decision variables assuming $(k,l)$ is transmitted.
\begin{equation}
U_{k,l}=|\sqrt{\mathcal{E}_{D,1,1}}+N_k|^2+|\sqrt{\mathcal{E}_{D,1,2}}+N_{l}|^2 \label{dv1}
\end{equation}
\begin{equation}
U_{k',l}=|N_{k'}|^2+|\sqrt{\mathcal{E}_{D,1,2}}+N_{l}|^2 \label{dv2}
\end{equation}
\begin{equation}
U_{k,l'}=|\sqrt{\mathcal{E}_{D,1,1}}+N_{k}|^2+|N_{l'}|^2 \label{dv3}
\end{equation}
\begin{equation}
U_{k',l'}=|N_{k'}|^2+|N_{l'}|^2 \label{dv4}
\end{equation}
where $U_{k,l}=|<\mathbf{Y}_1,\mathbf{S}_{m_{k}}>|^2+|<\mathbf{Y}_2,\mathbf{S}_{m_{l}}>|^2$. According to the decision variables from (\ref{dv1}) to (\ref{dv4}), the receiver chooses $(\hat{k},\hat{l})=\mathrm{argmax}_{\hat{k'}\hat{l'}}\;U_{k',l'}$ in the first round. An error is committed if any of the $U_{k',l}$, $U_{k,l'}$ and $U_{k',l'}$ is greater than $U_{k,l}$. 
The union bound on $P_{e}(k,l)$ is defined as 
$P_{e}(k,l)\leq \sum_{(k',l')\neq(k,l)}\Pr\left(u_{k,l}<u_{k',l'}|(k,l)\right)$.
In the following, we give the expression for each conditional probability where each decision variable given $(k,l)$ is transmitted in the first round. 
\begin{align} \label{eq:probdv1_2}
\Pr(U_{k,l}<U_{k',l'}|(k,l))&=\Pr(U_{k,l}<U_{k',l'}) \nonumber \\
&=\Pr(|\sqrt{\mathcal{E}_{D,1,1}}+N_k|^2+|\sqrt{\mathcal{E}_{D,1,2}}+N_{l}|^2<|N_{k'}|^2+|N_{l'}|^2)
\end{align}
\begin{align} \label{eq:probdv2}
\Pr(U_{k,l}<U_{k',l}|(k,l))&=\Pr(U_{k,l}<U_{k',l}) \nonumber \\
&=\Pr(|\sqrt{\mathcal{E}_{D,1,1}}+N_k|^2<|N_{k'}|^2)
\end{align}
\begin{align} \label{eq:probdv3_2}
\Pr(U_{k,l}<U_{k,l'}|(k,l))&=\Pr(U_{k,l}<U_{k,l'}) \nonumber \\
&=\Pr(|\sqrt{\mathcal{E}_{D,1,2}}+N_l|^2<|N_{l'}|^2)
\end{align}
In \cite[p. 686]{ProakisBook}, $P_{2}(L)$ is defined as the probability of error in choosing between $U_{k,l}$ and any other decision variable $U_{k',l}$, $U_{k,l'}$ or $U_{k',l'}$. 
Conditional probabilities for the second round given $(k,l)$ become cumulatively
\begin{align} \label{eq:probdv4}
\Pr(U_{k,l}<U_{k',l'}|(k,l))&=\Pr(U_{k,l}<U_{k',l'}) \nonumber \\
&=\Pr(|\sqrt{\mathcal{E}_{D,1,1}}+N_{k,1}|^2+|\sqrt{\mathcal{E}_{D,2,1}}+N_{k,2}|^2+|\sqrt{\mathcal{E}_{D,1,2}}+N_{l,1}|^2+|\sqrt{\mathcal{E}_{D,2,2}}+N_{l,2}|^2\nonumber \\
&<|N_{k',1}|^2+|N_{l',1}|^2+|N_{k',2}|^2+|N_{l',2}|^2)
\end{align}
\begin{align} \label{eq:probdv5}
\Pr(U_{k,l}<U_{k',l}|(k,l))&=\Pr(U_{k,l}<U_{k',l}) \nonumber \\
&=\Pr(|\sqrt{\mathcal{E}_{D,1,1}}+N_{k,1}|^2+|\sqrt{\mathcal{E}_{D,2,1}}+N_{k,2}|^2<|N_{k',1}|^2+|N_{k',2}|^2)
\end{align}
\begin{align} \label{eq:probdv6}
\Pr(U_{k,l}<U_{k,l'}|(k,l))&=\Pr(U_{k,l}<U_{k,l'}) \nonumber \\
&=\Pr(|\sqrt{\mathcal{E}_{D,1,2}}+N_{l,1}|^2+|\sqrt{\mathcal{E}_{D,2,2}}+N_{l,2}|^2<|N_{l',1}|^2+|N_{l',2}|^2)
\end{align}
The probabilities of one and both of the sources to be in error will be derived using the conditional probabilities in \ref{eq:probdv1_2}-\ref{eq:probdv6} in the upcoming subsections for the two different source distributions.

\subsection{Uniform Sources \label{sec:uni-dual}}

The first source $U_{1}$ is defined to be uniformly distributed over $(-\sqrt{3},\sqrt{3})$ and the second source $U_{2}$ is defined as $U_{2}=\rho U_{1}+\sqrt{1-\rho^2}U_{2}'$ based on $U_{1}$ and an auxiliary random vector $U_{2}'$ which is also uniform on $(-\sqrt{3},\sqrt{3})$. Depending on the value of $\rho$, the distribution of the second source $U_{2}$ can be either a triangular distribution or a contaminated uniform distribution. In the case of a high correlation, i.e. $\rho$ is very close to $1$, the effect of the auxiliary random variable $U_{2}'$ will be very small. On the contrary, for a low correlation between $U_{1}$ and $U_{2}$, $U_{2}'$ will have a significant effect so the second source will have a triangular distribution as a sum of the two uniform random vectors. We will focus on the extreme case of a very high correlation between the two sources. So, here we have one uniform and one almost uniform (contaminated uniform) source having covariance equal to the correlation coefficient $\rho$ between them.

\begin{figure}[htp]
  \centering
  \includegraphics[width=0.70\linewidth]{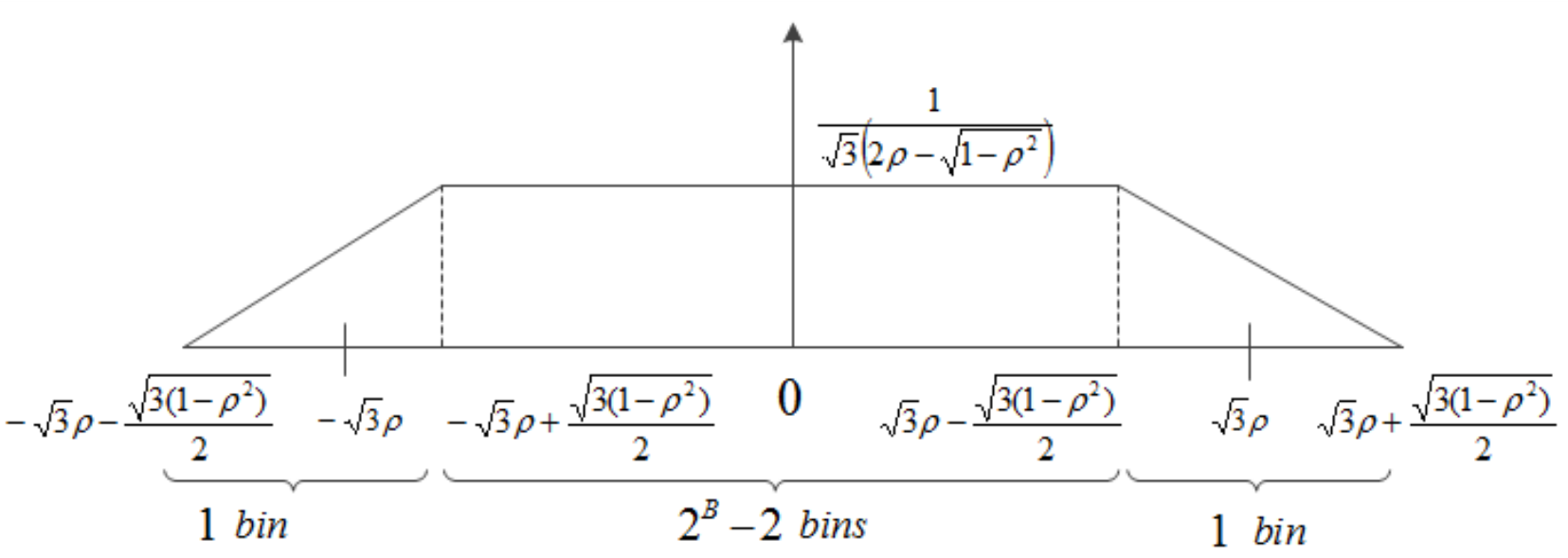}
  \caption{Pictorial representation of quantization process for the defined distribution with the allocation of the quantization bins}
	\label{fig:quantizer}
\end{figure}

The source messages are quantized as depicted in Figure \ref{fig:quantizer}, where each tail of the distribution is considered as one quantization bin and the interior part, which is composed by the remaining $2^B-2$ bins, is uniformly quantized. Note that for a full correlation between the sources, i.e. $\rho=1$,  the 'contamination' in the source distribution vanishes and the shape given by Figure \ref{fig:quantizer} becomes a rectangular.

At the end of the second round, the protocol is terminated with distortion bounded as
\begin{equation}
D=D_{q}(1-P_{e})+D_{e}P_{e} \leq D_{q}+D_{e,1}P_{e,1}+D_{e,2}P_{e,2} \label{distortion}
\end{equation}
where $P_{e}$ is the total probability of error which consists of $P_{e,1}$ and $P_{e,2}$ indicating the probability of error on one of the sources and both sources, respectively (for detailed derivation see (\ref{probability}) in Appendix \ref{sec:poe_both}). 
Both probabilities include the uncorrectable error in the first round. $D_{q}$ represents the distortion caused by the quantization process and $D_{e}$ corresponds to the MSE distortion for the case where an error was made. Splitting the distortion for the erroneous case, where $D_{e,1}$ denotes the distortion for one source in error and in the same way $D_{e,2}$ denotes the case when both sources incorrectly decoded. Let us denote the estimation error by $e$, so that its variance $E [u-\hat{u}|l\;in\;error]^2$ for $l=0$ yields the quantization distortion given by
\begin{equation} \label{dist_q}
D_q\leq (2^B-2)^{-2}\left(12+\frac{1-\rho^2}{\rho^2}-\frac{4\sqrt{3(1-\rho^2)}}{\rho}\right)+\frac{{3(1-\rho^2)}^{3/2}}{8\rho^3}.
\end{equation} 
$D_{e,1}$ is defined and bounded as follows
\begin{align} \label{dist_one}
D_{e,1}& = E \left[\left( u_{m}-\hat{u}_{m}\right)^2|u_m\;in\;error\right] \nonumber \\
&\leq 6\left(2^{-2B+2}+5(1-\rho^2)+2^{-B+3}\sqrt{1-\rho^2}\right)
\end{align} for the $m^{th}$ source where $m=1,2$. Note that for $m=2$ above given expression (\ref{dist_one}) becomes an equality. Finally, for the worst case when both sources are in error we have the following expansion and it is bounded as given by
\begin{align} \label{dist-two}
D_{e,2}&=\sum_{m=1}^{2} E \left[\left( u_{m}-\hat{u}_{m}\right)^2|u_m\;in\;error\right]\nonumber \\
&\leq 14+12\rho^2+3(1-\rho^2)/4+6\rho \sqrt{1-\rho^2}
\end{align}
$P_{e,1}$ is defined by 
\begin{equation} \label{eq:pe_1_uni}
P_{e,1}=\left\lceil 2^{B}\sqrt{1-\rho^2} \right\rceil \Pr(E_{e \rightarrow c,1})P_2(1)+\left\lceil 2^{B}\theta \sqrt{1-\rho^2} \right\rceil P_2(2)
\end{equation} through setting $P_2(1)$ for (\ref{eq:probdv2}), (\ref{eq:probdv3_2}) and $P_2(2)$ for (\ref{eq:probdv5}) and (\ref{eq:probdv6}). On the other hand for the case where both sources to be in error at the end of the first or the second round, the probability of error is achieved through setting $P_2(2)$ for (\ref{eq:probdv1_2}) and $P_2(4)$ for (\ref{eq:probdv4}).
\begin{equation} \label{eq:pe_2_uni}
P_{e,2}=\left\lceil 2^{B}\sqrt{1-\rho^2} \right\rceil 2^{B} \Pr(E_{e \rightarrow c,1})^2 P_2(2)+\left\lceil 2^{B}\theta \sqrt{1-\rho^2} \right\rceil 2^{B} P_2(4)
\end{equation}
Further detail on the derivation of the error probabilities (\ref{eq:pe_1_uni}) and (\ref{eq:pe_2_uni}) can be found in Appendix \ref{sec:poe_both}.
Through combining (\ref{eq:pe_1_uni}), (\ref{eq:pe_2_uni}),(\ref{dist_q}), (\ref{dist_one}), (\ref{dist-two}) with (\ref{distortion}), we get the following bound on distortion as
\begin{multline} \label{multi-distbound}
D\leq K_{1}D_q+\left(K_{2}\sqrt{1-\rho^2}e^{B\ln2}+K_3 \epsilon(\rho)\right)e^{(B-3)\ln2-\frac{\mathcal{E}_{\mathrm{D},1}+\mathcal{E}_{\mathrm{C},1}(\sqrt{\lambda}-1)^2}{2N_{0}}}D_{e,2} 
\\
+\left(K_{4}\sqrt{1-\rho^2}e^{B\ln2}+ K_{5}\epsilon(\rho)\right)e^{-\frac{\mathcal{E}_{\mathrm{D},1}+2\mathcal{E}_{\mathrm{C},1}(\sqrt{\lambda}-1)^2}{4N_{0}}}D_{e,1} 
\\
+\left(K_{6}\sqrt{1-\rho^2}e^{B\ln2}+ K_{7}\epsilon(\rho)\right)e^{(B-7)\ln2-\frac{\mathcal{E}_{\mathrm{D},1}+\mathcal{E}_{\mathrm{D},2}}{2N_{0}}} D_{e,2}
\\
+\left(K_{8}\sqrt{1-\rho^2}e^{B\ln2}+ K_{9}\epsilon(\rho)\right)e^{-\frac{\mathcal{E}_{\mathrm{D},1}+\mathcal{E}_{\mathrm{D},2}}{4N_{0}}} D_{e,1} 
\end{multline}
where $K_{1},K_{4},K_{5}$ are $O(1)$, $K_{2},K_{3}$ are $O(\mathcal{E}_{\mathrm{D},1})$, $K_{6},K_{7},K_{8},K_{9}$ are $O((\mathcal{E}_{\mathrm{D},1}+ \mathcal{E}_{\mathrm{D},2})^3)$ with $\epsilon(\rho)\in[0,1)$ which arose from the ceiling functions in (\ref{eq:pe_1_uni}) and (\ref{eq:pe_2_uni}). 

For a high level of correlation between the sources, i.e. when $\sqrt{1-\rho^2}<\theta 2^{-B}$, we set the relations of the energies as $\mathcal{E}_{C,1}=\frac{\mathcal{E}_{D,2}}{(1-\sqrt{\lambda})^{2}}$ and $\mathcal{E}_{D,2}=(2-\mu)\mathcal{E}_{D,1}$ where $\mu$ is an arbitrary constant satisfying $\mu\in(0,2)$. And the asymptotic bound for a high correlation level becomes
\begin{equation}\label{dist-highcorr}
D_{high}\leq e^{-\frac{\mathcal{E}_{\mathrm{D},1}(1-\mu/3)}{N_0}}\beta(\mathcal{E}_{\mathrm{D},1},\rho)
\end{equation}
where $$\beta(\mathcal{E}_{\mathrm{D},1},\rho)=\left(\frac{96+\frac{3}{\rho^2}e^{-\frac{\mathcal{E}_{\mathrm{D},1}}{2N_0}}}{14+\left(\frac{1}{2}e^{-\frac{\mathcal{E}_{\mathrm{D},1}}{2N_0}}+2\rho^2\right)^2}\right)^{2/3}$$ which arose from the distortion terms together with the ceiling functions.%$K_{h}=K_3+K_7$. 
To simplify the calculations the energy used by a source on a particular phase is assumed to be half of the energy on the corresponding round, e.g. $\mathcal{E}_{D,1}=2\mathcal{E}_{D,1,1}=2\mathcal{E}_{D,1,2}$.  Note that the exponential behaviour observed in (\ref{dist-highcorr}) is the same as a single source yields in \cite{EURECOM+3798} which is studied in detail in Section \ref{sec:Novel-feedback}. Furthermore, there is a difference of factor $1/2$ between the exponentials of (\ref{dist-highcorr}) and the information theoretic bounds (\ref{ineq:asym_hc_m}), (\ref{ineq:asym_uni_lc_m}) and (\ref{ineq:asympd3}).

The average energy $\mathcal{E}$ used by the protocol given by (\ref{eq:energy}) can be made arbitrarily close to $\mathcal{E}_{\mathrm{D},1}$ with 
vanishing $P_{e,1}$ and $P_{e,2}$, guaranteed by the interval in which $\epsilon(\rho)$ is defined.

\subsection{Gaussian Sources \label{sec:gauss-dual}} 

\begin{figure}[htp]
  \centering
  \includegraphics[width=0.70\linewidth]{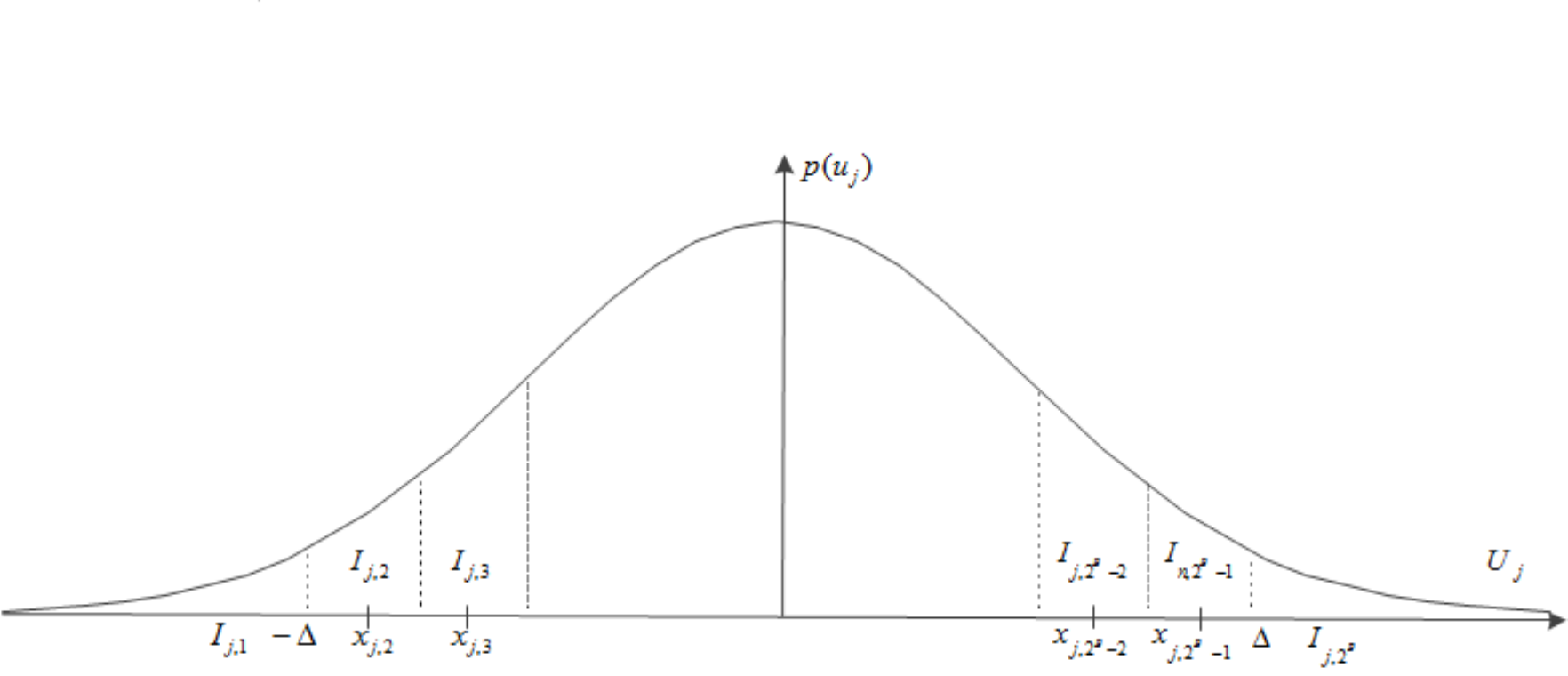}
  \caption{Linear quantization of $U_{j}$}
	\label{fig:quantization}
\end{figure}

Same structure of the sources from the uniform-contaminated uniform case is adapted to dual gaussian sources defined as in (\ref{eq:sourcers}) where $U_{1}$ and $U_{2}'$ are  normally distributed with zero mean and unit variance. Here $U_{2}'$ is used as an auxiliary random variable to define the relationship between the two sources $U_{1}$ and $U_{2}$ with the joint probability density function given below
\begin{equation}
f(u_{1},u_{2})=\frac{1}{2\pi\sqrt{1-\rho^2}}\exp\left[{-\frac{u_{1}^2-2\rho u_{1}u_{2}+u_{2}^2}{2(1-\rho^2)}}\right]
\end{equation}
for $-\infty<u_{1}<\infty$ and $\infty<u_{2}<\infty$. The definition of $U_{2}$ ensures that the covariance between the sources equals the correlation coefficient $\rho$. As in the uniform case, protocol can go up to two rounds each of which consists of two phases; a data phase and a control phase. 
The messages $m_{1}$ and $m_{2}$ will be discretized through uniform quantization, i.e. the bins are located equidistantly from each other and for each source the reconstruction points $x_{j,n}$ are the midpoints of the intervals $I_{j,n}$ which define each of the bins for the $j^{th}$ source with $n=2,...,2^{B}-1$. The quantization intervals corresponding to the tails of the bell curve ($I_{j,1}$ and $I_{j,2^{B}}$ for $j=1,2$) are considered as one bin for each side as shown in Figure \ref{fig:quantization}. 
The rest of the partitioning is made for each source as
\begin{equation}
I_{j,n}=[-\Delta+\frac{\Delta(n)}{2^{B-1}-1},-\Delta+\frac{\Delta(n+1)}{2^{B-1}-1}[,
\end{equation}
%\begin{equation}
%I_{2,j}=[-\Delta_{2}+\frac{\Delta_{2}j}{2^{B_{2}-1}-1},-\Delta_{2}+\frac{\Delta_{2}(j+1)}{2^{B_{2}-1}-1}[
%\end{equation}
with $\Delta=2\sqrt{B\ln 2}$. Let us set the quantization levels for each source as $x_{j,1}=-\Delta$ and $x_{j,2^{B}}=\Delta$. 
Unlike the scheme studied in the previous section, here the notion of compatible pairs arises from the statistical differences of normal distribution. $(m,n)$ is called a compatible pair if $|\rho U_{1}-U_{2}|<\theta$ for $\forall u_{1}, u_{2}\in B$ where $\theta$ is an arbitrary constant. This definition assures that, during the quantization process, the correlation between the two sources would not allow the second source to fall in a bin further than a certain distance. %So that the probability of $j$ given $i$ is bounded.
$J_{m}$ represents the set that $n$ is assumed to be contained. Outside of this set, the pair $(m,n)$ becomes incompatible with the corresponding probability of error $(1-\Pr(|U'_{2}|<\theta \sqrt{1-\rho^2}))$. In this case, the probability of having an error can be composed by three different events; both sources to be detected wrong, $\hat{u}_1$ detected correctly as $\hat{u}_2$ detected wrong or vice versa. These three events are summarized in two cases as only one source to be in error or both. 
The overall distortion at the end of the second round is defined and bounded by
\begin{align} \label{gauss-dist2}
D&=D_{q}(1-P_{e})+D_{e}P_{e}\nonumber \\
&\leq D_q+(1-\Pr(|U_2'|>\theta \sqrt{1-\rho^2}))\left(D_{e,c,1} P_{e,c,1}+D_{e,c,2} P_{e,c,2}\right)+\Pr(|U_2'|>\theta \sqrt{1-\rho^2})D_{e,ic,1}P_{e,ic,1}\nonumber \\
&+\Pr(|U_2'|>\theta \sqrt{1-\rho^2})D_{e,ic,2} P_{e,ic,2}\nonumber \\
&\overset{(a)}{\leq}D_q+D_{e,c,1} P_{e,c,1}+D_{e,c,2} P_{e,c,2}+\Pr(|U_2'|>\theta \sqrt{1-\rho^2})\left(D_{e,ic,1}+D_{e,ic,2} P_{e,ic,2}\right)
\end{align} where $ic$ and $c$ in the subscripts represent the incompatible and compatible pairs, respectively. $P_{e,ic,j}$ is the error probability of $j$ incompatible sources being in error whereas $P_{e,c,j}$ represents the probability of those which are compatible. $D_{e,ic,j}$ and $D_{e,c,j}$ denote the corresponding distortions for each case, respectively. Note that, error probabilities and the corresponding distortion levels for the case of both sources being in error are assumed to be equivalent, i.e. $P_{e,c,2}=P_{e,ic,2}=P_{e,2}$ and $D_{e,c,2}=D_{e,ic,2}=D_{e,2}$. It should be also noted that the probability of error only one incompatible source to be in error is upper bounded by 1. The derivations of the distortion terms for each case is given in detail in Appendix \ref{sec:dist-gauss}.
$P_{e,1}$ and $P_{e,2}$  are defined by
\begin{equation} \label{eq:pe_1}
P_{e,1}=\left\lceil 2^{B}\theta \sqrt{1-\rho^2} \right\rceil \Pr(E_{e \rightarrow c,1})P_2(1,\frac{\mathcal{E}_{\mathrm{D},1}}{2})+\left\lceil 2^{B}\theta \sqrt{1-\rho^2} \right\rceil P_2(2,\frac{\mathcal{E}_{\mathrm{D},1}+\mathcal{E}_{\mathrm{D},2}}{2})
\end{equation}
\begin{equation} \label{eq:pe_2}
P_{e,2}=\left\lceil 2^{B}\theta \sqrt{1-\rho^2} \right\rceil 2^{B} \Pr(E_{e \rightarrow c,1})^2 P_2(2,\mathcal{E}_{\mathrm{D},1})+\left\lceil 2^{B}\theta \sqrt{1-\rho^2} \right\rceil 2^{B} P_2(4,\mathcal{E}_{\mathrm{D},1}+\mathcal{E}_{\mathrm{D},2})
\end{equation}
where
\begin{equation} \label{eq:P_2(L)}
P_{2}(L,\gamma)= \frac{1}{2^{2L-1}} e^{-\gamma}\sum_{n=0}^{L-1} \left(\frac{1}{n!}\sum_{k=0}^{L-1-n}\dbinom{2L-1}{k}\right) \gamma^{n} \nonumber \\
\end{equation} in round $L$ given by the formula \cite[eq:12.1-24]{ProakisBook}. $\Pr(E_{e \rightarrow c,1})$, error probability of an uncorrectable error to occur in the first round, as defined and bounded in the previous section \ref{sec:uni-dual} and $\gamma$ represents the SNR.  Explicitly, in the first round for only one source being in error, the error probability is obtained by $P_2(1)$ whereas $P_2(2)$ gives the probability for both sources being in error. Accordingly $P_2(2)$ and $P_2(4)$ represent the probabilities in the second round.

The distortion level at the end of the second round (\ref{gauss-dist2}) is obtained by substituting error probabilities (\ref{eq:pe_1}) and (\ref{eq:pe_2}) with corresponding distortion terms derived in the Appendix \ref{sec:dist-gauss} into (\ref{gauss-dist2}) and given in the following explicit form as
\begin{align} \label{multi-distbound-g}
&D\leq K_{1}D_q+ K_{2} D_{e,ic,1} e^{-\frac{\theta^2 (1-\rho^2)}{2}}+\left(K_{3}\theta \sqrt{1-\rho^2} e^{B\ln2}+K_{4} \epsilon(\rho)\right)D_{e,2} e^{(B-3)\ln2-\frac{\mathcal{E}_{\mathrm{D},1}+\mathcal{E}_{\mathrm{C},1}(\sqrt{\lambda}-1)^2}{2N_{0}}}\nonumber \\
&+\left(K_{5}\theta \sqrt{1-\rho^2} e^{B\ln2}+ K_{6}\epsilon(\rho)\right)D_{e,c,1} e^{-\frac{\mathcal{E}_{\mathrm{D},1}+2\mathcal{E}_{\mathrm{C},1}(\sqrt{\lambda}-1)^2}{4N_{0}}}\nonumber \\
&+\left(K_{7}\theta \sqrt{1-\rho^2} e^{B\ln2}+ K_{8}\epsilon(\rho)\right)D_{e,2} e^{(B-7)\ln2-\frac{\mathcal{E}_{\mathrm{D},1}+\mathcal{E}_{\mathrm{D},2}}{2N_{0}}}\nonumber \\
&+\left(K_{9}\theta \sqrt{1-\rho^2} e^{B\ln2}+ K_{10}\epsilon(\rho)\right)D_{e,c,1} e^{-\frac{\mathcal{E}_{\mathrm{D},1}+\mathcal{E}_{\mathrm{D},2}}{4N_{0}}}\nonumber \\
&+\left(K_{11}\theta \sqrt{1-\rho^2} e^{B\ln2}+K_{12} \epsilon(\rho)\right)D_{e,2} e^{B\ln2-\frac{\theta^2(1-\rho^2)}{2}-\frac{\mathcal{E}_{\mathrm{D},1}+\mathcal{E}_{\mathrm{C},1}(\sqrt{\lambda}-1)^2}{2N_{0}}}\nonumber \\
&+\left(K_{13}\theta \sqrt{1-\rho^2} e^{B\ln2}+K_{14} \epsilon(\rho)\right)D_{e,2} e^{B\ln2-\frac{\theta^2(1-\rho^2)}{2}-\frac{\mathcal{E}_{\mathrm{D},1}+\mathcal{E}_{\mathrm{D},2}}{2N_{0}}}\nonumber \\
\end{align}
where $K_{2}=1/2$, $K_{3},K_{4},K_{5},K_{6},K_{11}$ and $,K_{12}$ are $O(\mathcal{E}_{\mathrm{D},1})$ and the rest of the factors are $O((\mathcal{E}_{\mathrm{D},1}+ \mathcal{E}_{\mathrm{D},2})^3)$ with $\epsilon(\rho)\in[0,1)$. % which arose from the ceiling functions in (\ref{jointprob-g}) and (\ref{totalprob-g}). 
For simplification in calculations, the energy used by a source on a particular phase is assumed to be half of the energy on the corresponding round, e.g. $\mathcal{E}_{D,1}=2\mathcal{E}_{D,1,1}=2\mathcal{E}_{D,1,2}$. 
Equating the order of the exponentials for the case of low correlation, i.e. $\theta>2\sqrt{\frac{B\ln2}{(1-\rho^2)}}$, we can set the relations of the energies as $\mathcal{E}_{\mathrm{C},1}=\frac{\mathcal{E}_{\mathrm{D},2}}{2(\sqrt{\lambda}-1)^2}$ and $\mathcal{E}_{D,2}=(2-\mu)\mathcal{E}_{D,1}$ where $\mu$ is an arbitrary constant within the interval $(0,2)$. 
\begin{equation}\label{dist-lowcorr-g}
D_{low}\leq e^{-\frac{\mathcal{E}_{\mathrm{D},1}(1-\mu/4)}{2N_0}}\gamma(\mathcal{E}_{\mathrm{D},1},\rho)+e^{-\frac{\mathcal{E}_{\mathrm{D},1}(1-\mu/3)}{2N_0}}\delta(\mathcal{E}_{\mathrm{D},1},\rho)+e^{-\frac{\mathcal{E}_{\mathrm{D},1}(3-\mu)}{4N_0}}\vartheta(\mathcal{E}_{\mathrm{D},1},\rho)
\end{equation}
where $\gamma$, $\omega$ and $\vartheta$ are functions of $\mathcal{E}_{\mathrm{D},1}$ and $\rho$ and arose from $K_3,K_4$, $K_5,K_6,K_9,K_10$ and $K_7,K_8$, respectively. For the case of high correlation, we set the relations of the energies as $\mathcal{E}_{C,1}=\frac{\mathcal{E}_{D,2}}{(1-\sqrt{\lambda})^{2}}$ and $\mathcal{E}_{D,2}=(2-\mu)\mathcal{E}_{D,1}$ where $\mu$ is an arbitrary constant satisfying $\mu\in(0,2)$ and the final bound becomes 
\begin{equation}\label{dist-highcorr-g}
D_{high}\leq e^{-\frac{\mathcal{E}_{\mathrm{D},1}(1-\mu/3)}{N_0}}\alpha(\mathcal{E}_{\mathrm{D},1})+K_6 e^{-\frac{\mathcal{E}_{\mathrm{D},1}(9-2\mu)}{4N_0}}+K_{10}e^{-\frac{\mathcal{E}_{\mathrm{D},1}(7-\mu)}{4N_0}} 
%+K_{high,2}e^{-\frac{\mathcal{E}_{\mathrm{D},1}(3-\mu)}{N_0}}+\alpha(\mathcal{E}_{\mathrm{D},1},\rho)
\end{equation} where $\alpha$ is a function of $\mathcal{E}_{\mathrm{D},1}$ which arose from $K_4$, $K_8$, $K_{12}$, $K_{14}$ together with the distortion terms %in the order of $(\mathcal{E}_{\mathrm{D},1})^{2/3}$. Explicitly 
and given by $\alpha(\mathcal{E}_{\mathrm{D},1},\rho)=\left(4\sqrt{\frac{\mathcal{E}_{\mathrm{D},1}}{\pi N_0}}+16\frac{\mathcal{E}_{\mathrm{D},1}}{N_0}\right)^{-2/3}$. The argument about the average energy used by the protocol made in uniform/contaminated uniform version is also applicable to Gaussian construction. The amount of energy used by the protocol is arbitrarily close to the energy consumed by the first data phase assured by vanishing error probability in this round.

The two extremes considered here (\ref{dist-lowcorr-g}) and (\ref{dist-highcorr-g}) show the effect of correlation on the reconstruction fidelity at the receiver.  
The high correlation case yields the exponential behaviour of the single-source case and benefits from energy accumulation,
or the collaboration of the two sources. Low-correlation results insignificantly reduced energy-efficiency.  In a large
network scenario, nodes with highly-correlated samples (in the above sense) would collaborate through joint detection at the 
receiver in order to optimize the energy efficiency of the network. 

%% file: 8_numerical.tex
\section{Numerical Results \label{sec:numerical}}

\begin{figure}
\centering
\includegraphics[width=.60\linewidth]{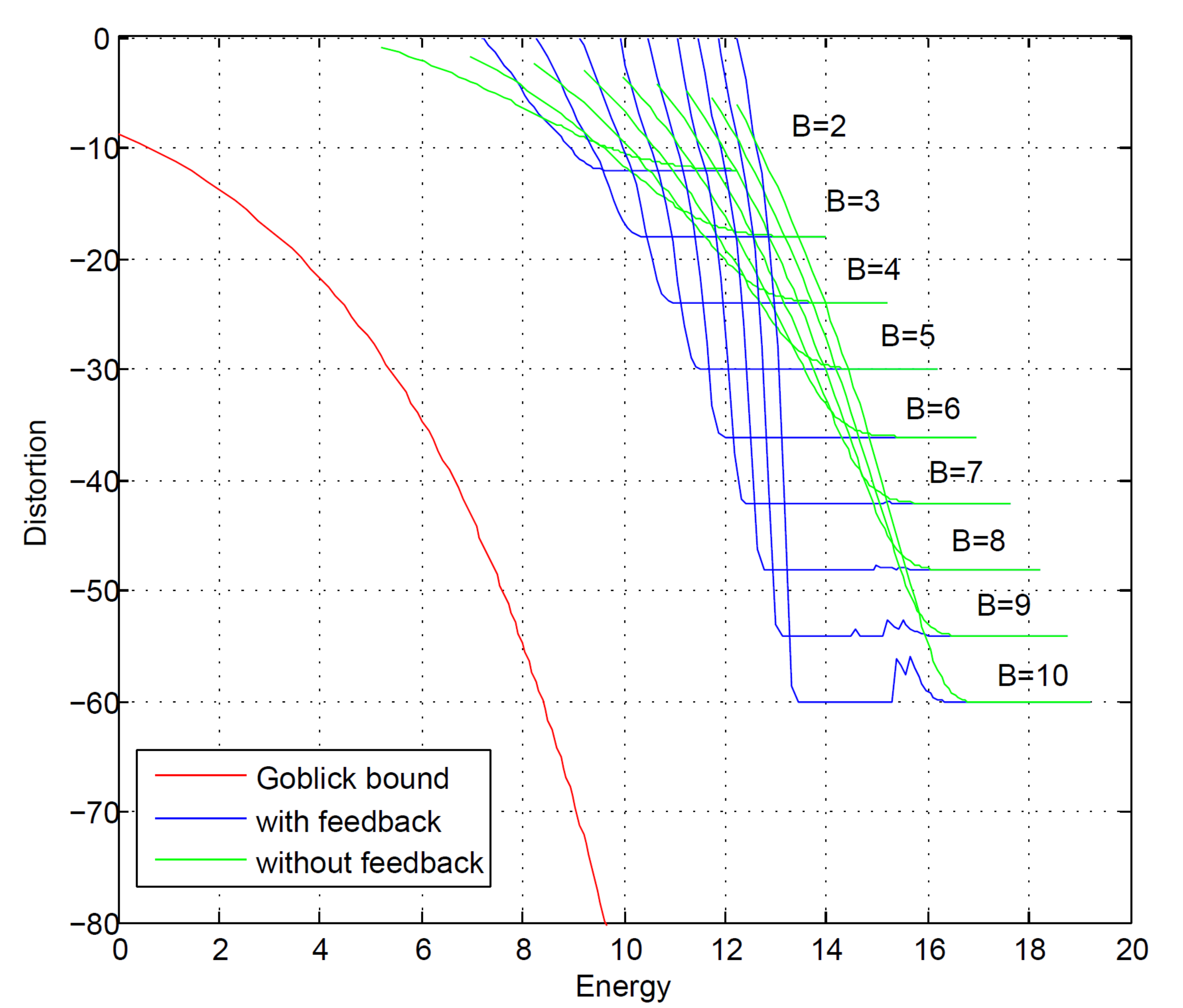}
\caption{Numerical evaluation of the derived bound on distortion for different values of $B$ in an AWGN channel.}
\label{fig:numeric1}
\end{figure}

In this section, we provide numerical evaluations of the bounds in (\ref{eq:distbound}) and (\ref{distortion}) for different values of $B$ and two rounds. In Figure \ref{fig:numeric1} we see the effect of going on for a second round on the distortion level subject to the average energy used by the protocol. The latter clearly provides an improvement in terms of distortion (approximately 3dB in energy efficiency).  Moreover, we see the predicted gap in energy-efficiency with respect to the outer-bound with a known channel. Furthermore, numerical analysis also confirmed the precision of the asymptotic result given in Section \ref{subsec:tworounds} by (\ref{finaldist}) regarding the relationship between the energies used in different rounds and phases.

Consider a simple wireless channel model instead of the AWGN channel studied in Section \ref{sec:Novel-feedback} where the channel amplitude and phase correspond to that of a Ricean channel with a ratio of the non-line-of-sight amplitude total signal amplitude $\alpha$.  In this case the output signal (\ref{eq:y_d}) becomes
\begin{equation}
\mathbf{Y'}_d = \sqrt{\mathcal{E}_{\mathrm{D},i}} \left(\sqrt{(1-\alpha)}\mathrm{e}^{j\Phi_i}+\sqrt{\alpha}\mathrm{h_{i}}\right) \mathbf{S}_m + \mathbf{Z}
\end{equation}
where $\mathrm{h_{i}}\sim N_{\Complex}(0,1)$ 
and $\alpha$ is in the range $[0,1]$. In this case only the statistics of the misdetected acknowledged error event is unchanged and is as given by (\ref{eq:prob_c_e}). The probability of an uncorrectable error becomes
\begin{align}
\Pr(E_{\mathrm{e}\rightarrow\mathrm{c},i}) &= \Pr\left(|\sqrt{(1-\alpha)\mathcal{E}_{\mathrm{C},i}}+\sqrt{\alpha \mathcal{E}_{\mathrm{C},i}} + z_c|^2 \leq \lambda\mathcal{E}_{\mathrm{C},i}\right)\nonumber\\
&=1-\mathrm{Q}_1\left(\sqrt{\frac{2(1-\alpha)\mathcal{E}_{\mathrm{C},i}}{\alpha \mathcal{E}_{\mathrm{C},i}+N_0}},\sqrt{\frac{2\lambda(1-\alpha)\mathcal{E}_{\mathrm{C},i}}{\alpha \mathcal{E}_{\mathrm{C},i}+N_0}}\right).
%&\overset{(a)}{\leq}1/2 \exp \left(-\frac{(\sqrt{\lambda}-1)^2 (1-\alpha)\mathcal{E}_{\mathrm{C},i}}{\alpha \mathcal{E}_{\mathrm{C},i}+N_0}\right) 
\label{eq:MQfunc_2}
\end{align}
The error probabilities $Pr(E_1)$ and $Pr(E_2)$ corresponding to the first and second rounds, respectively are derived using an adaptation of \cite[eq:12.1-22]{ProakisBook}, which is given by
\begin{equation} \label{eq:exactpe}
P_M= 1-\int_{0}^{\infty}\left(1-e^{-v(1+\alpha \gamma)}\sum_{k=0}^{L-1}\frac{(v(1+\alpha \gamma))^k}{k!}\right)^{M-1}\left[v\left(\frac{1+\alpha \gamma}{\gamma(1-\alpha)}\right)\right]^{\frac{L-1}{2}} e^{-v-\frac{\gamma(1-\alpha)}{(1+\alpha \gamma)}}I_{L-1}\left(2\sqrt{\frac{v\gamma(1-\alpha)}{1+\alpha \gamma}}\right)
\end{equation}
through numerical evaluation for $L=j$ for the $j^{th}$ round where $I_{L-1}$ is the modified bessel function of order $L-1$, $v=\frac{u}{2\mathcal{E}(N_0+\alpha \mathcal{E})}$ and $\gamma=\mathcal{E}/N_0$. $u$ is the first decision variable with a non-central chi-square distribution having $2L$ degrees of freedom and non-centrality parameter $s^2=\mathcal{E}^2(1-\alpha)$. Note that above probability is equivalent to its original version in \cite[eq:12.1-22]{ProakisBook} for $\alpha=0$. In the fading channel case, the protocol provides a more significant improvement when going from one to two rounds, due to the added diversity.  Here it should be expected that the use of more than two rounds could be even more beneficial, unlike the AWGN case.

\begin{figure}
\centering
\includegraphics[width=.60\linewidth]{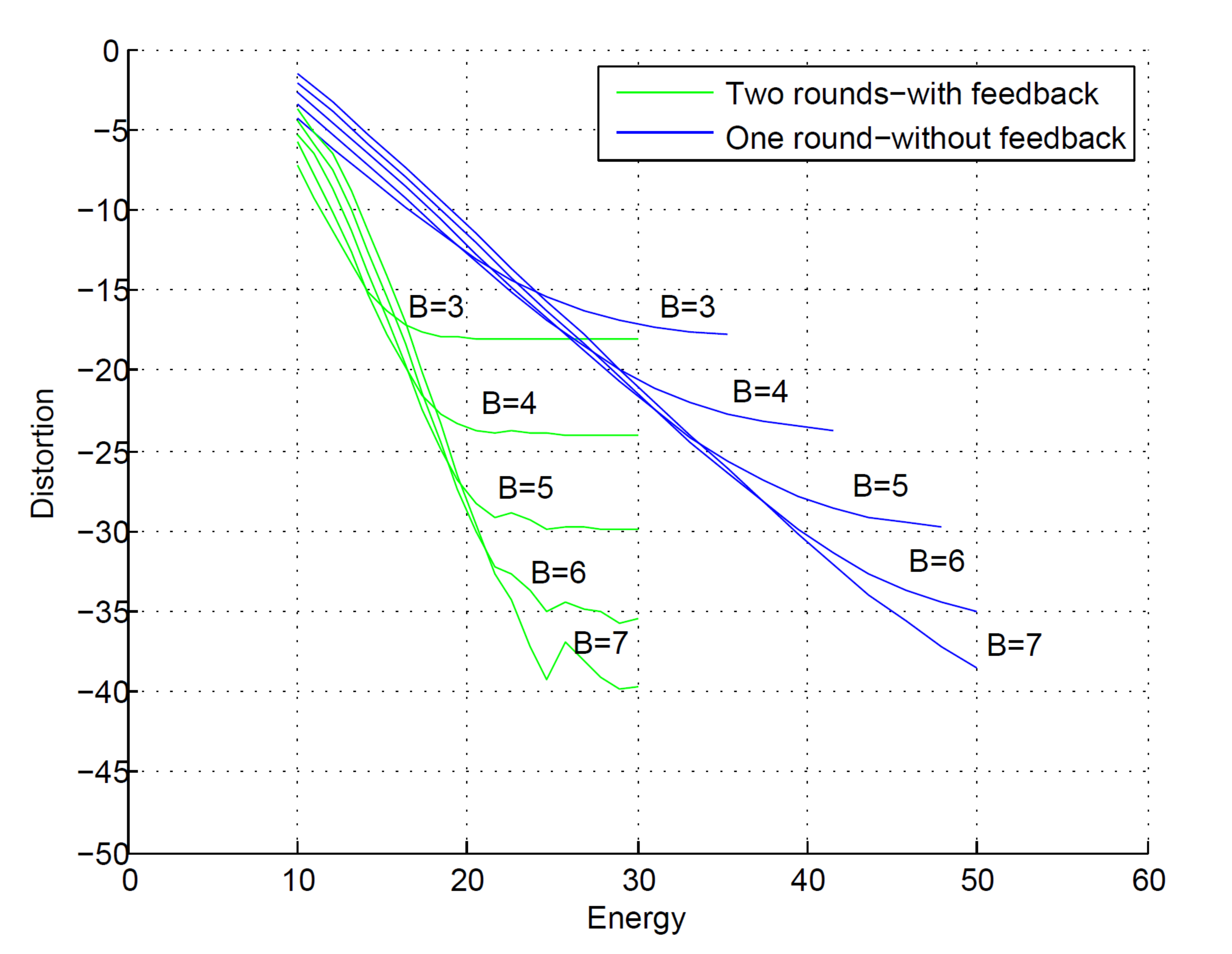}
\caption{Numerical evaluation of the distortion for different values of $B$ in a wireless channel for $\alpha=0.5$}
\label{fig:numeric3}
\end{figure}
\begin{figure}
\centering
\includegraphics[width=.60\linewidth]{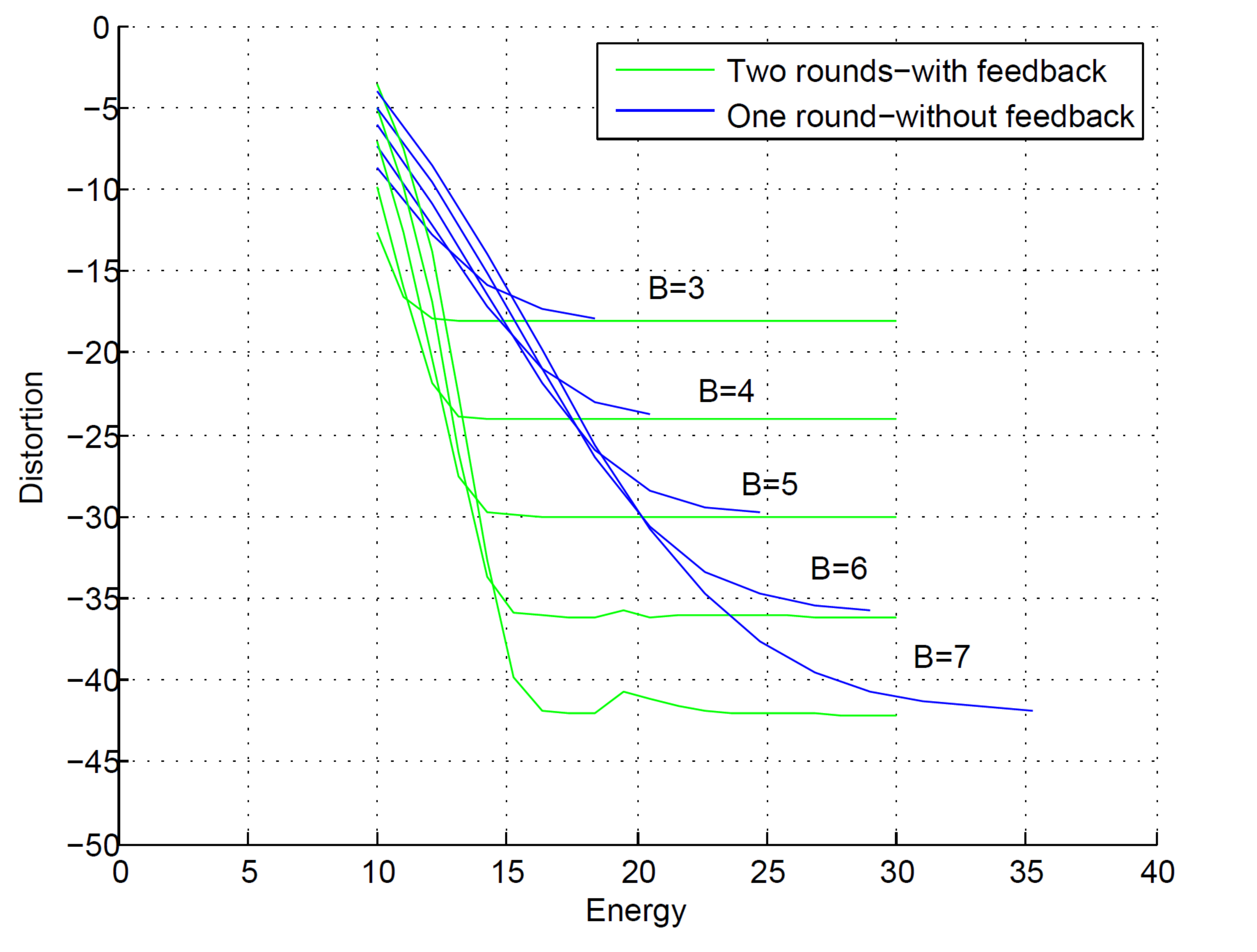}
\caption{Numerical evaluation of the distortion for different values of $B$ in a wireless channel for $\alpha=0.1$}
\label{fig:numeric4}
\end{figure}

The upper bound on the reconstruction error given in Section (\ref{subsec:tworounds}) by (\ref{eq:dist}) is adapted current case and by substituting (\ref{eq:exactpe}) and (\ref{eq:MQfunc_2}) we obtain the following bound on the distortion at the end of the second round.
\begin{align} \label{dist_w}
D\left(\mathcal{E},N_0,N,\lambda\right)&=2^{-2B}(1-P_e) + 2P_{e} \nonumber \\
&\leq 2^{-2B} +2 \left[P_M(L=1)\Pr(E_{\mathrm{e}\rightarrow\mathrm{c},1})+P_M(L=2)\right]
\end{align}
The change in upper bound (\ref{dist_w}) is depicted in Figure \ref{fig:numeric3} and Figure \ref{fig:numeric4} through numerical evaluation based on several $B$ values for the case $\alpha=0.5$ and $\alpha=0.1$, respectively.

\begin{figure}
\centering
\includegraphics[width=.60\linewidth]{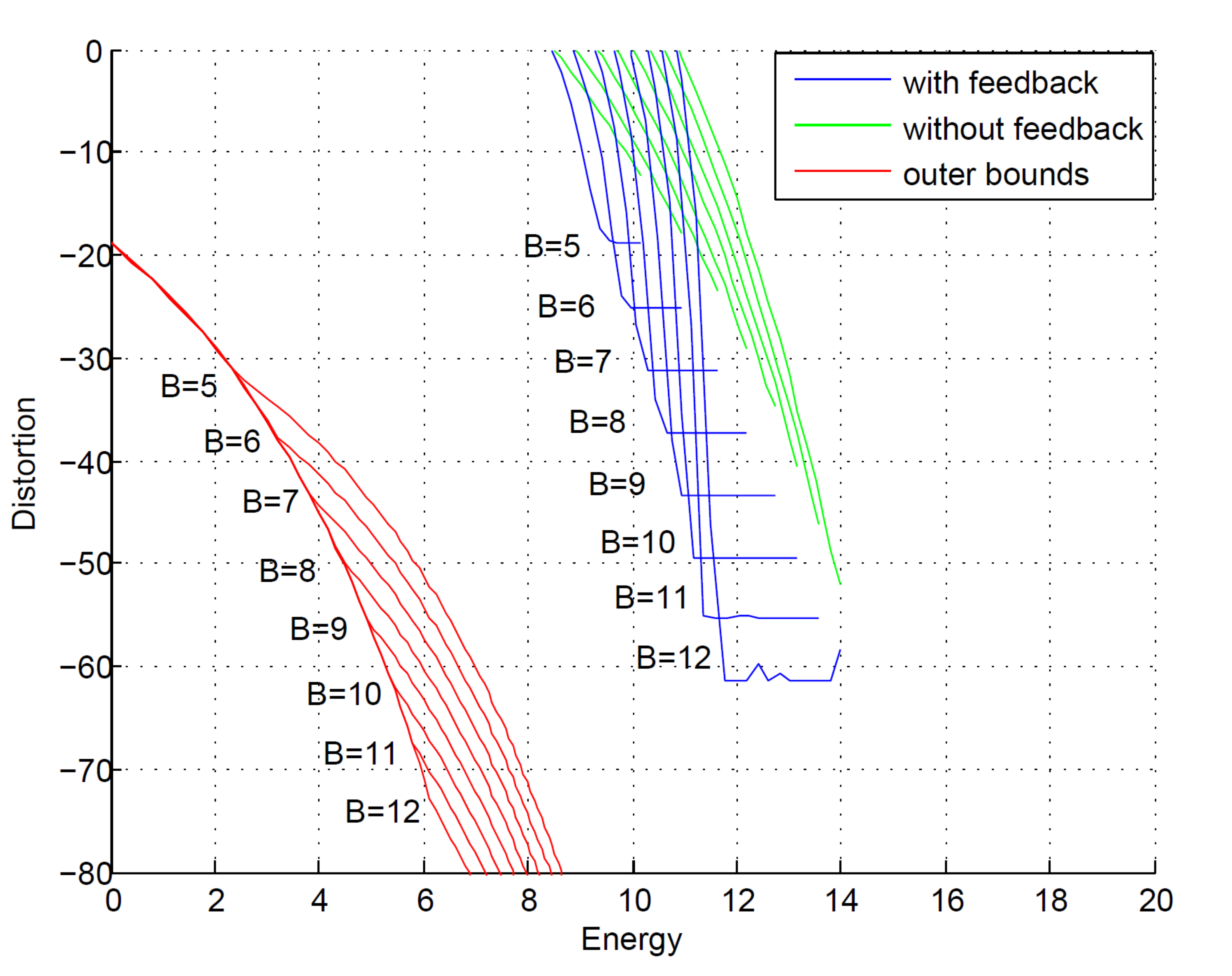}
\caption{Numerical evaluation of the derived upper and lower bounds on distortion for different values of $B$ for uniform/contaminated uniform dual-source case.}
\label{fig:numeric2}
\end{figure}

The numerical evaluation of the distortion bounds for the dual-source case are given by Figure \ref{fig:numeric2}. In this plot, the red curves represent the outer bounds (\ref{ineq:maxd2_u2}) derived in Section \ref{sec:sumch} for different $B$ values, where we have chosen $1-\rho^2 = 2^{-2B}$. The blue curves are the upper bound (\ref{distortion}) on distortion analyzed in Section \ref{sec:uni-dual}. The green curves are drawn for a protocol terminated after the first round which is the case without feedback.  We see from the lower-bounds that the energy accumulation remains feasible even at distortions below that of a uniform quantizer with $B$-bits (the asymptotes of the proposed sceheme). In practice, this suggests that the quantizer bin size should be chosen such that the difference in amplitude between the two sources should be on the order of the quantization error (i.e. 1-bit deviation between the sources).  We also see that the asymptotic performance does not emerge for small values of $B$ using the derived bounds, necessitating further numerical study of the proposed scheme in this case in order to better judge the gap from the lower-bounds. Nevertheless, the improvement using feedback is very significant, even for small values of $B$.

%% file: 9_conclusion.tex
\section{Conclusion \label{sec:conc}}

We derived lower bounds on the reconstruction error for the transmission of two correlated analog sources in the presence
of causal feedback. The bounds are specialized to the case of wideband channels. All our derivations are applied to two multiple-access channel types,
a sum channel and parallel channel and both for uniform and Gaussian sources. We obtain improvement with respect to the performance
achieved in [1] in terms of the asymptotic behaviour of the derived bounds on distortion with additional
feedback. 
We then introduce a low-latency two-way protocol for the transmission of a single random variable over a wideband channel and analyze its asymptotic behaviour with non-coherent 
detection for both uniform and Gaussian distributions. Another point worths mentioning is the
discussion made in Section II-A regarding to the comparison
between the performance of a single source and two highly
correlated sources. We show that the transmission of two highly correlated sources
can achieve the energy-efficiency of a single source with the same total energy, at least in certain regimes governing the level of correlation. Lastly, we find based on the results obtained in Section \ref{sec:third_r} that the gap between the outer bounds and the bounds obtained by our proposed feedback scheme cannot be closed (i.e. an improvement in terms of the asymptotic performance cannot be achieved) by repeating the protocol more than two rounds. This is supported by the numerical results provided for the single-source case.  We also present some performance examples of the proposed protocol on random fading channels.

%% file: 10_Appendix.tex
\section{Appendix} \label{sec:app}

\subsection{Appendix I \label{sec:discussion}}
Hereafter, we give the derivation of $I(\mathbf{U};\mathbf{\hat{U}})$ in two different expansions in order to bound the reconstruction error, first of which is based on the sources and given  as
\begin{align} \label{mutinf9}
I(\mathbf{U};\mathbf{\hat{U}})&=h(\mathbf{U})-h(\mathbf{U}|\mathbf{\hat{U}}) \nonumber \\
&\geq h(\mathbf{U})-h(\mathbf{U}-\mathbf{\hat{U}})
\end{align}
where $h(\mathbf{U})=\frac{K}{2}\log 2 \pi e$ for a normally distributed source dimension of $K$ and $h(\mathbf{U})=K\log 2\sqrt{3}$ for a uniform source.
\begin{align}
h(\mathbf{U}-\mathbf{\hat{U}})&=\sum_{i=1}^{K}h(U_{i}-\hat{U}_{i})\nonumber \\
&\leq \sum_{i=1}^{K}\frac{1}{2}\log(\mathbb{E}[(U_{i}-\hat{U}_{i})^2])\nonumber \\
&\leq \frac{K}{2}\log(2\pi e D).
\end{align}
Secondly, the same mutual information is expanded based on the output signals as follows
\begin{align} \label{mutinf10}
I(\mathbf{U};\mathbf{\hat{U}})&\leq I(\mathbf{X}_{1},\mathbf{X}_{2};\mathbf{Y}|\mathbf{\Phi}) \nonumber \\
&=h(\mathbf{Y}|\mathbf{\Phi})-h(\mathbf{Y}|\mathbf{X}_{1},\mathbf{X}_{2},\mathbf{\Phi}) \nonumber \\
&\leq \sum_{i=1}^{N}h(Y_{i}|\mathbf{\Phi})-h(\mathbf{Z}) \nonumber \\
&\leq N\left(\sum_{i=1}^{N}\log(\mathbb{E}[Y_{i}^2])-\log(NN_0)\right) \nonumber \\
&=N\log \left(1+\frac{K\mathcal{E}}{NN_0}\right)
\end{align} which is applicable to both distributions.
%Equating the two different expressions of $I(\mathbf{U};\mathbf{\hat{U}})$ yields the lower bound (\ref{ineq:asympd3}).

\subsection{Appendix II-Sum Channel High Correlation \label{sec:app_sum_high}}
The mutual information $I(\mathbf{U}_{m};\mathbf{Y})$ is derived through two different expansions where the first expansion is
\begin{align}
I(\mathbf{U}_{m};\mathbf{Y})&\leq I(\mathbf{U}_{m};\mathbf{Y},\Phi_{m},\Phi_{m'}) \nonumber \\
%&=I(\mathbf{U}_{m};\mathbf{Y}|\Phi_{m},\Phi_{m'})+I(\mathbf{U}_{m};\Phi_{m},\Phi_{m'})\nonumber \\
&=h(\mathbf{Y}|\Phi_{m},\Phi_{m'})-h(\mathbf{Y}|\mathbf{U}_{m},\Phi_{m},\Phi_{m'}) \nonumber \\
&=\sum_{i=1}^{N} h(Y_{i}|Y^{i-1},\Phi_{m},\Phi_{m'})- \sum_{i=1}^{N} h(Y_{i}|Y^{i-1},\mathbf{U}_{m},\Phi_{m},\Phi_{m'}) \nonumber \\
&\leq \sum_{i=1}^{N} h(Y_{i}|Y^{i-1},\Phi_{m},\Phi_{m'})- \sum_{i=1}^{N} h(Y_{i}|Y^{i-1},\mathbf{U}_{m},\mathbf{X}_{m}e^{j\phi_{m}},\mathbf{X}_{m'}e^{i\phi_{m'}},\Phi_{m},\Phi_{m'}) \nonumber \\
&=\sum_{i=1}^{N} h(Y_{i}|Y^{i-1},\Phi_{m},\Phi_{m'})- \sum_{i=1}^{N} h(Z_{i}) \nonumber \\
&\leq \sum_{i=1}^{N} \log \left(1+\frac{\mathcal{E}_{m,i}+\mathcal{E}_{m',i}}{N N_0}\right) \nonumber \\
&\leq N\log \left(1+\frac{\sum_{i=1}^{N}(\mathcal{E}_{m,i}+\mathcal{E}_{m',i})}{N N_0}\right) \nonumber \\
&\leq N\log \left(1+\frac{K(\mathcal{E}_{m,i}+\mathcal{E}_{m',i})}{N N_0}\right). \label{ineq:A}
\end{align} 
The expansion of $I(\mathbf{U}_{m};\mathbf{Y})$ given above is independent of the distribution and the source number, which means that (\ref{ineq:A}) applies also for the source $m'$ and both of the uniform and Gaussian cases. 
On the other hand, for the second expansion of the same mutual information we have
\begin{align} \label{ineq:B}
I(\mathbf{U}_{m};\mathbf{Y})&=h(\mathbf{U}_{m})-h(\mathbf{U}_{m}-\mathbf{\hat{U}}_{m}|\mathbf{Y}) \nonumber \\ 
&\geq h(\mathbf{U}_{m})-h(\mathbf{U}_{m}-\mathbf{\hat{U}}_{m}).
\end{align}
The required entropies in the uniform-contaminated uniform case for $m=1,2$, we have
\begin{equation} \label{ineq:C1}
h(\mathbf{U}_{1})=K \log 2\sqrt{3}, 
\end{equation}
\begin{align}\label{ineq:C2}
h(\mathbf{U}_{2})&=h(\rho\mathbf{U}_{1}+\sqrt{1-\rho^2}\mathbf{U}_{2}')\nonumber \\
&\geq \frac{K}{2}\log \left(2^{\frac{2}{K}(K\log|\rho|+h(\mathbf{U}_{1})}+2^{\frac{2}{K}h(\sqrt{1-\rho^2}\mathbf{U}_{2}'}\right)\nonumber \\
&=\frac{K}{2}\log \left(2^{\frac{2}{K}K\log|\rho|2\sqrt{3}}+2^{\frac{2}{K}K\log 2\sqrt{3|1-\rho^2|}}\right)\nonumber \\
%&=\frac{K}{2}\log \left(|\rho|2\sqrt{3}+2\sqrt{3|1-\rho^2|}\right).
&=K \log 2\sqrt{3}.
\end{align}
For the Gaussian case the entropies of the two sources are given by
\begin{equation} \label{ineq:C3}
h(\mathbf{U}_{m})=\frac{K}{2}\log 2\pi e.
\end{equation} 
The final term required to derive the second expansion of (\ref{ineq:B}) is given by
\begin{align} \label{ineq:C4}
h(\mathbf{U}_{m}-\mathbf{\hat{U}}_{m})&\leq \sum_{j=1}^{K}h(U_{m,j}-\hat{U}_{m,j}) \nonumber \\
&\leq \frac{K}{2}\log \left(2\pi e \frac{1}{K}\sum_{j=1}^{K}\mathbb{E}[(U_{m,j}-\hat{U}_{m,j})^2]\right) \nonumber \\
&\leq K\log \left(\sqrt{2\pi e D_{m}}\right)
\end{align}
which is independent of the distribution of the sources. Substituting (\ref{ineq:C1}) and (\ref{ineq:C4}) for $m=1$ into (\ref{ineq:B}), yields the second expansion of the desired mutual information for the first source. In the same way, (\ref{ineq:C1}) and (\ref{ineq:C4}) with $m=2$ is substituted into (\ref{ineq:B}).

%\subsection{Appendix III- Sum Channel High Correlation for Source Model II  \label{sec:app_sum_highII}}

%The difference between the two source models effects the second expansion through the entropy of a source for the uniform case, since the second model brings along symmetry. Thus the two expansions (\ref{ineq:A}) and (\ref{ineq:B}) of $I(\mathbf{U}_{m};\mathbf{Y})$ apply to source model II. But for the uniform version, we get $h(\mathbf{U}_{m})\geq K \log \left(|\rho|2\sqrt{3}+2\sqrt{3|1-\rho^2|}\right)$ for $m=1,2$, which yields
%\begin{equation} \label{ineq:AII}
%I(\mathbf{U}_{m};\mathbf{Y})\geq K\log \left(\frac{2\sqrt{3}(\rho+\sqrt{1-\rho^2})}{\sqrt{2\pi e D_m}}\right).
%\end{equation}
%Equating the two expansions (\ref{ineq:A}) and (\ref{ineq:AII}) of the same mutual information $I(\mathbf{U}_{m};\mathbf{Y})$ gives us the lower bound on the reconstruction error in estimating $\mathbf{U}_{m}$ for the uniform case. As noted before, the two source models provide an identical expression for normally distributed sources.

\subsection{Appendix III- Sum Channel Low Correlation  \label{sec:app_sum_low}}
First expansion is
\begin{align}
&I(\mathbf{U}_{m};\mathbf{Y}|\mathbf{U}_{m'})\leq I(\mathbf{U}_{m};\mathbf{Y}|\mathbf{U}_{m'},\Phi_{m},\Phi_{m'})\nonumber \\
&=h(\mathbf{Y}|\mathbf{U}_{m'},\Phi_{m},\Phi_{m'})-h(\mathbf{Y}|\mathbf{U}_{m},\mathbf{U}_{m'},\Phi_{m},\Phi_{m'}) \nonumber \\
&=\sum_{i=1}^{N} h(Y_{i}|Y^{i-1},\mathbf{U}_{m'},\Phi_{m},\Phi_{m'})-\sum_{i=1}^{N} h(Y_{i}|Y^{i-1},\mathbf{U}_{m},\mathbf{U}_{m'},\Phi_{m},\Phi_{m'})\nonumber \\
&=\sum_{i=1}^{N} h(Y_{i}|Y^{i-1},\mathbf{U}_{m'},\mathbf{X}_{m'}e^{i\phi_{m'}},\Phi_{m},\Phi_{m'})-\sum_{i=1}^{N} h(Y_{i}|Y^{i-1},\mathbf{U}_{m},\mathbf{U}_{m'},\mathbf{X}_{m}e^{j\phi_{m}},\mathbf{X}_{m'}e^{i\phi_{m'}},\Phi_{m},\Phi_{m'})\nonumber \\
&\overset{(a)}{=}\sum_{i=1}^{N} h(X_{m,i}e^{j\phi_{m,i}}+Z_{i}|Y^{i-1},\mathbf{U}_{m'},\Phi_{m},\Phi_{m'})-\sum_{i=1}^{N} h(Z_{i}) \nonumber \\
&\leq \sum_{i=1}^{N} h(X_{m,i}e^{j\phi_{m,i}}+Z_{i})-\sum_{i=1}^{N} h(Z_{i})\nonumber \\
&\leq N \log \left(\sum_{i=1}^{N}\log(Var(X_{m,i}e^{j\phi_{m,i}}+Z_{i}))-\log (Var(\mathbf{Z}))\right) \nonumber \\
&=N\log \left(1+\frac{K \mathcal{E}_{m}}{N N_{0}}\right). \label{ineq:F}
\end{align}
In step (a) $\mathbf{X}_{m'}e^{i\phi_{m'}}$ is subtracted from the output signal, which provides $\mathbf{X}_{m}e^{j\phi_{m}}$ together with the noise term in the next step. For the second expansion based on the sources, we have
\begin{align}
I(\mathbf{U}_{m};\mathbf{Y}|\mathbf{U}_{m'})&= h(\mathbf{U}_{m}|\mathbf{U}_{m'})-h(\mathbf{U}_{m}|\mathbf{U}_{m'},\mathbf{Y}) \nonumber \\
&=h(\mathbf{U}_{m}|\mathbf{U}_{m'})-h(\mathbf{U}_{m}-\hat{\mathbf{U}}_{m}|\mathbf{U}_{m'},\mathbf{Y}) \nonumber \\
&\geq h(\mathbf{U}_{m}|\mathbf{U}_{m'})-h(\mathbf{U}_{m}-\hat{\mathbf{U}}_{m}). \label{ineq:G}
\end{align}
The conditional entropy $h(\mathbf{U}_{m}|\mathbf{U}_{m'})$ is obtained for $m=1,2$
\begin{align} \label{ineq:G1}
h(\mathbf{U}_{1}|\mathbf{U}_{2})&=-I(\mathbf{U}_{1};\mathbf{U}_{2})+h(\mathbf{U}_{1})\nonumber \\
&=-h(\mathbf{U}_{2})+h(\mathbf{U}_{2}|\mathbf{U}_{1})+h(\mathbf{U}_{1})\nonumber \\
&\overset{(a)}{\geq}-\frac{K}{2}\log 2\pi e+ K\log 2\sqrt{3|1-\rho^2|}+K\log 2\sqrt{3} \nonumber \\
&=K\log \left(\frac{12\sqrt{|1-\rho^2|}}{\sqrt{2\pi e}}\right)
\end{align}
\begin{align}\label{ineq:G2}
h(\mathbf{U}_{2}|\mathbf{U}_{1})&=h(\rho \mathbf{U}_{1}+\sqrt{1-\rho^2}\mathbf{U}_{2'}|\mathbf{U}_{1})\nonumber \\
%&=h(\mathbf{U}_{2}|\mathbf{U}_{1})\nonumber  \\
&=h(\sqrt{1-\rho^2}\mathbf{U}_{2'})\nonumber \\
&=K\log 2\sqrt{3|1-\rho^2|}
\end{align}
,respectively. In step (a) of (\ref{ineq:G1}), the entropy of $\mathbf{U}_{2}$ is bounded by the entropy of a standard gaussian random vector. 
For the gaussian distribution, the conditional entropy of one source given the other is obtained as 
\begin{align} \label{ineq:G3}
h(\mathbf{U}_{1}|\mathbf{U}_{2})&=-I(\mathbf{U}_{1};\mathbf{U}_{2})+h(\mathbf{U}_{1})\nonumber \\
&\overset{(a)}{=}-h(\mathbf{U}_{2})+h(\mathbf{U}_{2}|\mathbf{U}_{1})+h(\mathbf{U}_{1})\nonumber \\
&=h(\mathbf{U}_{2}|\mathbf{U}_{1})\nonumber \\
&=\frac{K}{2}\log (1-\rho^2)2\pi e
\end{align}
where in the step (a), we used the equality of the entropies between two standard normal random variables.

\subsection{Appendix IV- Bound on product Distortion $D_{p}$ for Sum Channel \label{sec:product}}

\label{subsec:productb}
The mutual information  $I(\mathbf U_{m},\mathbf U_{m'};\mathbf Y)$ is obtained as
 \begin{align} 
I(\mathbf U_{m},\mathbf U_{m'};\mathbf Y)&\leq I(\mathbf U_{m},\mathbf U_{m'};\mathbf Y|\boldsymbol\Phi) \nonumber \\
&=h(\mathbf{Y}|\mathbf{\Phi})-h(\mathbf{Y}|\mathbf{U}_{m},\mathbf{U}_{m'},\mathbf{\Phi}) \nonumber \\
&=h(\mathbf{Y}|\mathbf{\Phi})-\sum_{i=1}^{N}h(Y_{i}|Y^{i-1},\mathbf{U}_{m},\mathbf{U}_{m'},\mathbf{\Phi}) \nonumber \\
&\leq \sum_{i=1}^{N}h(Y_{i}|\mathbf{\Phi})-\sum_{i=1}^{N}h(Y_{i}|Y^{i-1},\mathbf{U}_{m},\mathbf{X}_{m},\mathbf{U}_{m'},\mathbf{X}_{m'},\mathbf{\Phi}) \nonumber \\
&=\sum_{i=1}^{N}h(Y_{i}|\mathbf{\Phi})-\sum_{i=1}^{N}h(Z_{i}).
\end{align} 
The variance of the received signal $Y_{i}$ becomes $\sum_{i=1}^{N}Var(Y_{i})=K(\mathcal{E}_{m}+\mathcal{E}_{m'})+NN_{0}$ and the desired mutual information is obtained as
 \begin{equation} \label{ineq:mutinf3}
I(\mathbf U_{m},\mathbf U_{m'};\mathbf Y|\boldsymbol\Phi)\leq N\log (1+\frac{K(\mathcal{E}_{m}+\mathcal{E}_{m'})}{NN_{0}}).
\end{equation} 
We also have for the uniform contaminated uniform construction the following expansion
\begin{align} \label{ineq:mutinf4}
I(\mathbf U_{m},\mathbf U_{m'};\mathbf Y)&\geq I(\mathbf{U}_{m},\mathbf{U}_{m'};\mathbf{\hat{U}}_{m},\mathbf{\hat{U}}_{m'})\nonumber \\
&\geq h(\mathbf{U}_{m},\mathbf{U}_{m'})-h(\mathbf{U}_{m}-\mathbf{\hat{U}}_{m})-h(\mathbf{U}_{m'}-\mathbf{\hat{U}}_{m'})\nonumber \\
&\geq \frac{K}{2}\log 144(1-\rho^2)-\frac{K}{2}\log(2\pi e)^{2}D_{p} \nonumber \\
&=\frac {K}{2}\log \left(\frac{36(1-\rho^2)}{\pi^2 e^2 D_{p}}\right)
\end{align} where $D_p = D_1 D_2$.
Two expressions (\ref{ineq:mutinf3}) and (\ref{ineq:mutinf4}) of the same mutual information are equated to obtain (\ref{ineq:product}). (\ref{ineq:mutinf4}) differs slightly for normally distributed sources as

\begin{align} \label{ineq:mutinf4gauss}
I(\mathbf U_{m},\mathbf U_{m'};\mathbf Y)&\geq I(\mathbf{U}_{m},\mathbf{U}_{m'};\mathbf{\hat{U}}_{m},\mathbf{\hat{U}}_{m'})\nonumber \\
&\geq h(\mathbf{U}_{m},\mathbf{U}_{m'})-h(\mathbf{U}_{m}-\mathbf{\hat{U}}_{m})-h(\mathbf{U}_{m'}-\mathbf{\hat{U}}_{m'})\nonumber \\
&\geq \frac{K}{2}\log (2\pi e)^2 (1-\rho^2)-\frac{K}{2}\log(2\pi e)^{2}D_{m}D_{m'} \nonumber \\
&=\frac {K}{2}\log \left(\frac{(1-\rho^2)}{ D_{p}}\right).
\end{align} 

\subsection{Appendix V- Parallel Channel High Correlation \label{sec:app_par_high}}
First expansion of the mutual information $I(\mathbf{U}_{m};\mathbf{Y}_{m},\mathbf{Y}_{m'})$ is
\begin{align}
&I(\mathbf{U}_{m};\mathbf{Y}_{m},\mathbf{Y}_{m'})\leq I(\mathbf{U}_{m};\mathbf{Y}_{m},\mathbf{Y}_{m'}|\Phi_{m},\Phi_{m'})\nonumber \\
&=h(\mathbf{Y}_{m},\mathbf{Y}_{m'}|\Phi_{m},\Phi_{m'})-h(\mathbf{Y}_{m},\mathbf{Y}_{m'}|\mathbf{U}_{m},\Phi_{m},\Phi_{m'})\nonumber \\
&\leq h(\mathbf{Y}_{m}|\Phi_{m},\Phi_{m'})+h(\mathbf{Y}_{m'}|\Phi_{m},\Phi_{m'})-h(\mathbf{Y}_{m}|\mathbf{U}_{m},\Phi_{m},\Phi_{m'})-h(\mathbf{Y}_{m'}|\mathbf{U}_{m},\mathbf{Y}_{m},\Phi_{m},\Phi_{m'})\nonumber \\
&= h(\mathbf{Y}_{m})+h(\mathbf{Y}_{m'})-h(\mathbf{Y}_{m}|\mathbf{U}_{m},\Phi_{m})-h(\mathbf{Y}_{m'}|\mathbf{U}_{m},\mathbf{Y}_{m},\Phi_{m'})\nonumber \\
&\leq h(\mathbf{Y}_{m})+h(\mathbf{Y}_{m'})-\sum_{i=1}^{N}h(\mathbf{Y}_{m,i}|\mathbf{Y}_{m}^{i-1},\mathbf{U}_{m},\mathbf{X}_{m}e^{i\phi_{m}},\Phi_{m}) -\sum_{i=1}^{N}h(\mathbf{Y}_{m',i}|\mathbf{Y}_{m'}^{i-1},\mathbf{U}_{m},\mathbf{Y}_{m},\mathbf{X}_{m'}e^{j\phi_{m'}},\Phi_{m'}) \nonumber \\
&\leq N\log \left(K\mathcal{E}_{m}+NN_{0}\right)+N\log\left(K\mathcal{E}_{m'}+NN_{0}\right)-N\log (NN_{0})^2 \nonumber \\
&=N\log \left(1+\frac{K\mathcal{E}_{m}}{NN_{0}}\right)\left(1+\frac{K\mathcal{E}_{m'}}{NN_{0}}\right). \label{ineq:L}
\end{align}
Second expansion which is based on the sources is as follows
\begin{align}
I(\mathbf{U}_{m};\mathbf{Y}_{m},\mathbf{Y}_{m'})&=h(\mathbf{U}_{m})-h(\mathbf{U}_{m}|\mathbf{Y}_{m},\mathbf{Y}_{m'})\nonumber \\
&=h(\mathbf{U}_{m})-h(\mathbf{U}_{m}-\mathbf{\hat{U}}_{m}|\mathbf{Y}_{m},\mathbf{Y}_{m'}) \nonumber \\
&\geq h(\mathbf{U}_{m})-h(\mathbf{U}_{m}-\mathbf{\hat{U}}_{m}) \label{ineq:M}
\end{align} for which, the entropy expressions introduced in Section \ref{sec:highcorrsumch} for both of the sources and the distribution types are substituted to achieve final form of the second expansion.

For the first source in uniform-contaminated uniform case, i.e. $m=1$, (\ref{ineq:C1}), for the second source (when $m=2$) (\ref{ineq:C2}) and finally for both $m=1,2$ in the Gaussian case (\ref{ineq:C3}) is used for the entropy $h(\mathbf{U}_{m})$. Since the entropy bound (\ref{ineq:C4}) applies to all possible scenarios, it is adapted in all of the cases considered. 

\subsection{Appendix VI- Parallel Channel Low Correlation \label{sec:app_par_low}}

Unlike the high correlation case, here the output signal cannot obtain information about one of the sources through the other one since benefiting from correlation is not possible. So the first expansion is given by
\begin{align}
I(\mathbf{U}_{m};\mathbf{Y}_{m}|\mathbf{U}_{m'},\mathbf{Y}_{m'})&\leq I(\mathbf{U}_{m};\mathbf{Y}_{m}|\mathbf{U}_{m'},\mathbf{Y}_{m'},\Phi_{m})\nonumber \\
&=h(\mathbf{Y}_{m}|\mathbf{U}_{m'},\mathbf{Y}_{m'},\Phi_{m})-h(\mathbf{Y}_{m}|\mathbf{U}_{m'},\mathbf{U}_{m},\mathbf{Y}_{m'},\Phi_{m}) \nonumber \\
&=\sum_{i=1}^{N} h(Y_{m,i}|Y_{m}^{i-1},\mathbf{U}_{m'},\mathbf{Y}_{m'},\Phi_{m})- \sum_{i=1}^{N} h(Y_{m,i}|Y_{m}^{i-1},\mathbf{U}_{m},\mathbf{U}_{m'},\mathbf{Y}_{m'},\Phi_{m}) \nonumber \\
&\leq \sum_{i=1}^{N} h(Y_{m,i}|Y_{m}^{i-1},\mathbf{U}_{m'},\mathbf{Y}_{m'},\mathbf{X}_{m'}e^{j\phi_{m'}},\Phi_{m})-\sum_{i=1}^{N} h(Y_{m,i}|Y_{m}^{i-1},\mathbf{U}_{m},\mathbf{U}_{m'},\mathbf{Y}_{m'},\mathbf{X}_{m}e^{i\phi_{m}},\Phi_{m}) \nonumber \\
& \leq N\log \left( 2 \pi e(K \mathcal{E}_{m}+N N_{0})\right)-N \log (2 \pi e N N_{0})\nonumber \\
&=N\log \left(1+\frac{K\mathcal{E}_{m}}{NN_{0}}\right) \label{ineq:O}
\end{align}
Second expansion is
\begin{align} \label{ineq:N}
I(\mathbf{U}_{m};\mathbf{Y}_{m}|\mathbf{U}_{m'},\mathbf{Y}_{m'})&=h(\mathbf{U}_{m}|\mathbf{U}_{m'},\mathbf{Y}_{m'})-h(\mathbf{U}_{m}|\mathbf{U}_{m'},\mathbf{Y}_{m},\mathbf{Y}_{m'}) \nonumber \\
&=h(\mathbf{U}_{m}|\mathbf{U}_{m'},\mathbf{Y}_{m'})-h(\mathbf{U}_{m}|\mathbf{U}_{m'},\mathbf{Y}_{m},\mathbf{Y}_{m'}) \nonumber \\
&\geq h(\mathbf{U}_{m}|\mathbf{U}_{m'})-h(\mathbf{U}_{m}-\hat{\mathbf{U}}_{m})\nonumber \\
\end{align}
which ends up in the identical form with the one given in the sum-channel low-correlation case showed by (\ref{ineq:G}). The conditional entropies $h(\mathbf{U}_{m}|\mathbf{U}_{m'})$ were derived for different combinations of the sources and the distributions already in Section \ref{sec:lowcorrsumch}.

\subsection{Appendix VII- Bound on product Distortion $D_{p}$ for Parallel Channel \label{sec:product_p}}

First expansion of the mutual information $I(\mathbf{U}_{m};\mathbf{Y}_{m},\mathbf{Y}_{m'})$ is obtained as
\begin{align}
&I(\mathbf{U}_{m},\mathbf{U}_{m'};\mathbf{Y}_{m},\mathbf{Y}_{m'})\leq I(\mathbf{U}_{m},\mathbf{U}_{m'};\mathbf{Y}_{m},\mathbf{Y}_{m'}|\Phi_{m},\Phi_{m'})\nonumber \\
&=h(\mathbf{Y}_{m},\mathbf{Y}_{m'}|\Phi_{m},\Phi_{m'})-h(\mathbf{Y}_{m},\mathbf{Y}_{m'}|\mathbf{U}_{m},\mathbf{U}_{m'},\Phi_{m},\Phi_{m'})\nonumber \\
&\leq h(\mathbf{Y}_{m}|\Phi_{m},\Phi_{m'})+h(\mathbf{Y}_{m'}|\Phi_{m},\Phi_{m'})-h(\mathbf{Y}_{m}|\mathbf{U}_{m},\mathbf{U}_{m'},\Phi_{m},\Phi_{m'})-h(\mathbf{Y}_{m'}|\mathbf{U}_{m},\mathbf{U}_{m'},\mathbf{Y}_{m},\Phi_{m},\Phi_{m'})\nonumber \\
&= h(\mathbf{Y}_{m})+h(\mathbf{Y}_{m'})-h(\mathbf{Y}_{m}|\mathbf{U}_{m},\mathbf{U}_{m'},\Phi_{m})-h(\mathbf{Y}_{m'}|\mathbf{U}_{m},\mathbf{U}_{m'},\Phi_{m'})\nonumber \\
&= h(\mathbf{Y}_{m})+h(\mathbf{Y}_{m'})-\sum_{i=1}^{N}h(\mathbf{Y}_{m,i}|\mathbf{Y}_{m}^{i-1},\mathbf{U}_{m},\mathbf{U}_{m'},\mathbf{X}_{m}e^{i\phi_{m}},\Phi_{m}) -\sum_{i=1}^{N}h(\mathbf{Y}_{m',i}|\mathbf{Y}_{m'}^{i-1},\mathbf{U}_{m},\mathbf{U}_{m'},\mathbf{X}_{m'}e^{j\phi_{m'}},\Phi_{m'}) \nonumber \\
&\leq N\log \left(K\mathcal{E}_{m}+NN_{0}\right)+N\log\left(K\mathcal{E}_{m'}+NN_{0}\right)-N\log (NN_{0})^2 \nonumber \\
&=N\log \left(1+\frac{K\mathcal{E}_{m}}{NN_{0}}\right)\left(1+\frac{K\mathcal{E}_{m'}}{NN_{0}}\right), \label{product_p1}
\end{align}

We also have for the uniform contaminated uniform construction the following expansion
\begin{align} \label{product_p2}
I(\mathbf U_{m},\mathbf U_{m'};\mathbf Y_{m},\mathbf Y_{m'} )&\geq I(\mathbf{U}_{m},\mathbf{U}_{m'};\mathbf{\hat{U}}_{m},\mathbf{\hat{U}}_{m'})\nonumber \\
&\geq h(\mathbf{U}_{m},\mathbf{U}_{m'})-h(\mathbf{U}_{m}-\mathbf{\hat{U}}_{m})-h(\mathbf{U}_{m'}-\mathbf{\hat{U}}_{m'})\nonumber \\
&\geq \frac{K}{2}\log 144(1-\rho^2)-\frac{K}{2}\log(2\pi e)^{2}D_{p} \nonumber \\
&=\frac {K}{2}\log \left(\frac{36(1-\rho^2)}{\pi^2 e^2 D_{p}}\right).
\end{align} where $D_p = D_1 D_2$.
Two expressions (\ref{product_p1}) and (\ref{product_p2}) of the same mutual information are equalized to obtain (\ref{ineq:product}). (\ref{ineq:mutinf4}) differs slightly as given in the following for normally distributed sources

\begin{align} \label{product_p3}
I(\mathbf U_{m},\mathbf U_{m'};\mathbf Y_{m},\mathbf Y_{m'})&\geq I(\mathbf{U}_{m},\mathbf{U}_{m'};\mathbf{\hat{U}}_{m},\mathbf{\hat{U}}_{m'})\nonumber \\
&\geq h(\mathbf{U}_{m},\mathbf{U}_{m'})-h(\mathbf{U}_{m}-\mathbf{\hat{U}}_{m})-h(\mathbf{U}_{m'}-\mathbf{\hat{U}}_{m'})\nonumber \\
&\geq \frac{K}{2}\log (2\pi e)^2 (1-\rho^2)-\frac{K}{2}\log(2\pi e)^{2}D_{m}D_{m'} \nonumber \\
&=\frac {K}{2}\log \left(\frac{(1-\rho^2)}{ D_{p}}\right).
\end{align} 

\subsection{Appendix VIII- Parallel Channel Alternative \label{sec:app_par_D2}}
\subsubsection{Uniform-Contaminated Uniform Case}
First expansion of $I(\mathbf{U}_{2};\mathbf{Y}_{2}|\mathbf{Y}_{1})$ is derived as
\begin{align} \label{ineq:mutinf12}
I(\mathbf{U}_{2};\mathbf{Y}_{2}|\mathbf{Y}_{1})&\leq I(\mathbf{U}_{2};\mathbf{Y}_{2}|\mathbf{Y}_{1},\Phi_{1},\Phi_{2})\nonumber \\
&=h(\mathbf{Y}_{2}|\mathbf{Y}_{1},\Phi_{2})-h(\mathbf{Y}_{2}|\mathbf{U}_{2},\mathbf{Y}_{1},\Phi_{2}) \nonumber \\
&=h(\mathbf{Y}_{2}|\mathbf{Y}_{1},\Phi_{2})-h(\mathbf{Y}_{2}|\mathbf{U}_{2},\mathbf{Y}_{1},\Phi_{2}) \nonumber \\
&\leq \sum_{i=1}^{N} h(Y_{2,i}|Y_{2}^{i-1},\Phi_{2})-\sum_{i=1}^{N} h(Y_{2,i}|Y_{2}^{i-1},\mathbf{U}_{2},\mathbf{X}_{2}e^{j\phi_{2}},\Phi_{2}) \nonumber \\
&=\sum_{i=1}^{N} h(Y_{2,i}|Y_{2}^{i-1},\Phi_{2})-\sum_{i=1}^{N} h(Z_{2,i}) \nonumber \\
&\leq \sum_{i=1}^{N} \log \left(1+\frac{\mathcal{E}_{2,i}}{NN_0}\right) \nonumber \\
&\leq N\log \left(1+\frac{\sum_{i=1}^{N}\mathcal{E}_{2,i}}{NN_0}\right) \nonumber \\
&\leq N\log(1+\frac{K\mathcal{E}_{2}}{NN_{0}})
\end{align}
Second expansion is given by
\begin{align}
I(\mathbf{U}_{2};\mathbf{Y}_{2}|\mathbf{Y}_{1})&=h(\mathbf{U}_{2}|\mathbf{Y}_{1})-h(\mathbf{U}_{2}|\mathbf{Y}_{2},\mathbf{Y}_{1}) \nonumber \\
&=h(\mathbf{U}_{2}|\mathbf{Y}_{1})-h(\mathbf{U}_{2}-\mathbf{\hat{U}}_{2}|\mathbf{Y}_{2},\mathbf{Y}_{1}) \nonumber \\
&=h(\rho \mathbf{U}_{1}+\sqrt{1-\rho^2}\mathbf{U}_{2}'|\mathbf{Y}_{1})-h(\mathbf{U}_{2}-\mathbf{\hat{U}}_{2}|\mathbf{Y}_{2},\mathbf{Y}_{1}) \nonumber \\
&\overset{(a)}{\geq} \frac{K}{2}\log \left(2^{\frac{2}{K}h(\sqrt{1-\rho^2}\mathbf{U}_{2}'|\mathbf{Y}_{1})}+2^{\frac{2}{K}(K\log|\rho|+h(\mathbf{U}_{1}|\mathbf{Y}_{1})}\right)-h(\mathbf{U}_{2}-\mathbf{\hat{U}}_{2})\nonumber \\
&=\frac{K}{2}\log \left(2^{\frac{2}{K}h(\sqrt{1-\rho^2}\mathbf{U}_{2}')}+2^{\frac{2}{K}(K\log|\rho|+h(\mathbf{U}_{1}|\mathbf{Y}_{1})}\right)-h(\mathbf{U}_{2}-\mathbf{\hat{U}}_{2}) \label{mutinf11}
\end{align}
here in step (a), we used the entropy-power inequality in order to expand the entropy $h(\rho \mathbf{U}_{1}+\sqrt{1-\rho^2}\mathbf{U}_{2}'|\mathbf{Y}_{1})$. And we obtain  $h(\mathbf{U}_{1}|\mathbf{Y}_{1})$in a general form as follows
\begin{align} \label{parso}
h(\mathbf{U}_{1}|\mathbf{Y}_{1})&=h(\mathbf{U}_{1})-I(\mathbf{U}_{1};\mathbf{Y}_{1})\nonumber \\
&\geq h(\mathbf{U}_{1})-I(\mathbf{U}_{1};\mathbf{Y}_{1}|\Phi_{1})\nonumber \\
&=h(\mathbf{U}_{1})-\left(h(\mathbf{Y}_{1}|\Phi_{1})-h(\mathbf{Y}_{1}|\Phi_{1},\mathbf{U}_{1})\right) \nonumber \\
&=h(\mathbf{U}_{1})-(h(\mathbf{Y}_{1}|\Phi_{1})-h(\mathbf{Y}_{1}|\mathbf{U}_{1},\Phi_{1}))\nonumber \\
&=h(\mathbf{U}_{1})-\sum_{i=1}^{N} h(Y_{1,i}|Y_{1}^{i-1},\Phi_{1})+\sum_{i=1}^{N} h(Y_{1,i}|Y_{1}^{i-1},\mathbf{U}_{1},\mathbf{X}_{1}e^{i\phi_{1}},\Phi_{1}) \nonumber \\
&=h(\mathbf{U}_{1})-\sum_{i=1}^{N} h(Y_{1,i}|Y_{1}^{i-1},\Phi_{1})+\sum_{i=1}^{N} h(Z_{1,i}) \nonumber \\
&\geq h(\mathbf{U}_{1})-\sum_{i=1}^{N} \log \left(1+\frac{\mathcal{E}_{1,i}}{NN_0}\right) \nonumber \\
&\geq h(\mathbf{U}_{1})-N\log \left(1+\frac{\sum_{i=1}^{N}\mathcal{E}_{1,i}}{NN_0}\right) \nonumber \\
&\geq h(\mathbf{U}_{1})-N\log \left(1+\frac{K\mathcal{E}_{1}}{NN_{0}}\right).
\end{align}

\subsubsection{Gaussian Case}

The first expansion  (\ref{ineq:mutinf12}) can be written in a general form to cover both of the sources as
\begin{equation}
I(\mathbf{U}_{m};\mathbf{Y}_{m}|\mathbf{Y}_{m'})\leq N\log(1+\frac{K\mathcal{E}_{m}}{NN_{0}}).
\end{equation}
For the second expansion, we have
\begin{align}
I(\mathbf{U}_{m};\mathbf{Y}_{m}|\mathbf{Y}_{m'})&=h(\mathbf{U}_{m}|\mathbf{Y}_{m'})-h(\mathbf{U}_{m}|\mathbf{Y}_{m},\mathbf{Y}_{m'}) \nonumber \\
&=h(\mathbf{U}_{m}|\mathbf{Y}_{m'})-h(\mathbf{U}_{m}-\mathbf{\hat{U}}_{m}|\mathbf{Y}_{m},\mathbf{Y}_{m'}) \nonumber \\
&\geq h(\mathbf{U}_{m}|\mathbf{Y}_{m'})-h(\mathbf{U}_{m}-\mathbf{\hat{U}}_{m})
\end{align}
For $m=1$
\begin{align}
h(\mathbf{U}_{1}|\mathbf{Y}_{2})&\geq \frac{K}{2}\log \left(2^{\frac{2}{K} h(\frac{1}{\rho}\mathbf{U}_{2}|\mathbf{Y}_{2})}+2^{\frac{2}{K}h(\frac{\sqrt{1-\rho^2}}{\rho}\mathbf{U'}_{2}|\mathbf{Y}_{2})}\right)\nonumber \\
&\overset{(a)}{\geq} \frac{K}{2}\log \left(2^{\frac{2}{K}h(\frac{1}{\rho}\mathbf{U}_{2}|\mathbf{Y}_{2})}+2^{\frac{2}{K}h(\frac{\sqrt{1-\rho^2}}{\rho}\mathbf{U'}_{2})}\right)\nonumber \\
&= \frac{K}{2} \log \left(\left(1+\frac{K\mathcal{E}_{2}}{NN_{0}}\right)^{-\frac{2N}{K}}\frac{2\pi e}{\rho^2}+\left(\frac{1-\rho^2}{\rho^2}\right)2\pi e\right)
\end{align} 
In step (a), the condition is neglected given that the output signal $\mathbf{Y}_{2}$ and auxiliary random vector $\mathbf{U'}_{2}$ is independent. Note that for the entropy of $\mathbf{U}_{1}$, we used $\mathbf{U}_{1}=\frac{1}{\rho} \mathbf{U}_{1}+\frac{\sqrt{1-\rho^2}}{\rho}\mathbf{U}_{2}'$. Consequently, second expansion for $m=1$ becomes
\begin{equation}
I(\mathbf{U}_{1};\mathbf{Y}_{1}|\mathbf{Y}_{2})\geq  \frac{K}{2} \log \frac{\left(\left(1+\frac{K\mathcal{E}_{2}}{NN_{0}}\right)^{-\frac{2N}{K}}\frac{1}{\rho^2}+\left(\frac{1-\rho^2}{\rho^2}\right)\right)}{D_1}
\end{equation}
For the second source, i.e. $m=2$, we have
\begin{equation}
I(\mathbf{U}_{2};\mathbf{Y}_{2}|\mathbf{Y}_{1})\geq  \frac{K}{2} \log \frac{\left(\left(1+\frac{K\mathcal{E}_{1}}{NN_{0}}\right)^{-\frac{2N}{K}}\rho^2+\left(1-\rho^2\right)\right)}{D_2}
\end{equation}
Equating the two expansions of $I(\mathbf{U}_{m};\mathbf{Y}_{m}|\mathbf{Y}_{m'})$ yields the lower bounds given in Section \ref{sec:tightpar}.

\subsection{Appendix IX- Probability of Error for the Achievable Scheme Dual-Source Case \label{sec:poe_both}}

The probability of the error is bounded by
\begin{align} \label{probability}
P_{e}&= \Pr(E_{1,1},E_{1,2}^c)\Pr(E_{e\rightarrow c,1,1})+\Pr(E_{1,1}^c,E_{1,2})\Pr(E_{e\rightarrow c,1,2})+\Pr(E_{1,1},E_{1,2})\Pr(E_{e\rightarrow c,1,1})\Pr(E_{e\rightarrow c,1,2}) \nonumber \\
&+(\Pr(E_{1,1},E_{1,2})(1-\Pr(E_{e\rightarrow c,1,1})\Pr(E_{e\rightarrow c,1,2})) \nonumber \\ 
&+\Pr(E_{1,1},E_{1,2}^c)(1-\Pr(E_{e\rightarrow c,1,1}))+\Pr(E_{1,1}^c,E_{1,2})(1-\Pr(E_{e\rightarrow c,1,2})))\Pr(E_{2}|E_{1}) \nonumber \\ 
&+(1-\Pr(E_{1,1},E_{1,2}))\Pr(E_{c\rightarrow e,1,1})\Pr(E_{c\rightarrow e,1,2})\Pr(E_{2}|E_{1}^c) \nonumber \\ 
&\overset{(a)}{=}\Pr(E_{e\rightarrow c,1,j})[\Pr(E_{1,1},E_{1,2}^c)+\Pr(E_{1,1}^c,E_{1,2})]+\Pr(E_{e\rightarrow c,1,j})^2 \Pr(E_{1,1},E_{1,2})\nonumber \\
&+\Pr(E_{2}|E_{1})[\Pr(E_{1,1},E_{1,2})(1-\Pr(E_{e\rightarrow c,1,j})^2)+(\Pr(E_{1,1},E_{1,2}^c)+\Pr(E_{1,1}^c,E_{1,2}))(1-\Pr(E_{e\rightarrow c,1,j}))]\nonumber \\
&+\Pr(E_{2}|E_{1}^c)[\Pr(E_{1,1}^c,E_{1,2}^c)\Pr(E_{c\rightarrow e,1,j})^2]\nonumber \\
&\overset{(b)}{\leq}\Pr(E_{e\rightarrow c,1,j})[\Pr(E_{1,1},E_{1,2}^c)+\Pr(E_{1,1}^c,E_{1,2})]+\Pr(E_{e\rightarrow c,1,j})^2 \Pr(E_{1,1},E_{1,2})\nonumber \\
&+\Pr(E_{2}|E_{1})[\Pr(E_{1,1},E_{1,2})+\Pr(E_{1,1},E_{1,2}^c)+\Pr(E_{1,1}^c,E_{1,2})]+\Pr(E_{2}|E_{1}^c)\Pr(E_{1,1}^c,E_{1,2}^c)\nonumber \\
&\overset{(c)}{=}\Pr(E_{e\rightarrow c,1,j})P_{e,1,1}+\Pr(E_{e\rightarrow c,1,j})^2 P_{e,2,1} + \Pr(E_{2})
\end{align} In step (a) the probability of an uncorrectable $\Pr(E_{e\rightarrow c,1,j})$ and misdetected acknowledged error $\Pr(E_{e\rightarrow c,1,j})$ are assumed to be equal for both sources whereas in (b) the probability of being decoded correctly, i.e. $(1-\Pr(E_{e\rightarrow c,1,j}))$, and the misdetection is upper bounded by 1. In the final step (c), the probability of only one source and both of the sources to be in error in the first round is denoted by $P_{e,1,1}$ and $ P_{e,2,1}$, respectively.

\subsection{Appendix X- Derivations of the Distortion terms \label{sec:dist-gauss}}

The distortion caused by quantization process $D_q$, by channel itself when both sources are in error $D_{e,2}$ are given by
\begin{align} \label{eq:distquant}
D_q&=\sum_{m=1}^{2^{B}}\sum_{n=1}^{2^{B}}\int_{I_{1,m}}\int_{I_{2,n}} \left[(u_{1}-\hat{u}_{1}(m))^2+(u_{2}-\hat{u}_{2}(n))^2\right]f(u_{1},u_{2})du_{2}du_{1}\nonumber \\
&=\sum_{m=1}^{2^{B}}\int_{I_{1,m}}(u_{1}-\hat{u}_{1}(m))^2 \sum_{n=1}^{2^{B}} \int_{I_{2,n}}f(u_{1},u_{2})du_{2}du_{1}+\sum_{n=1}^{2^{B}}\int_{I_{2,n}}(u_{2}-\hat{u}_{2}(n))^2 \sum_{m=1}^{2^{B}} \int_{I_{1,m}}f(u_{1},u_{2})du_{1}du_{2}\nonumber \\
&=\sum_{m=1}^{2^{B}}\int_{I_{1,m}}(u_{1}-\hat{u}_{1}(m))^2 f(u_{1})du_{1} + \sum_{n=1}^{2^{B}}\int_{I_{2,n}}(u_{2}-\hat{u}_{2}(n))^2 f(u_{2})du_{2} \nonumber \\
&=\int_{\Delta}^{\infty}(u_1-\Delta)^2 f(u_1)du_1+\int_{\Delta}^{\infty}(u_2-\Delta)^2 f(u_2)du_2+\int_{-\infty}^{-\Delta}(u_1+\Delta)^2 f(u_1)du_1+\int_{-\infty}^{-\Delta}(u_2+\Delta)^2 f(u_2)du_2 \nonumber \\
&+\sum_{m=2}^{2^{B}-1}\int_{I_{1,m}}(u_{1}-\hat{u}_{1}(m))^2 f(u_{1})du_{1} + \sum_{n=2}^{2^{B}-1}\int_{I_{2,n}}(u_{2}-\hat{u}_{2}(n))^2 f(u_{2})du_{2} \nonumber \\
&\leq 4\left(e^{-\Delta^2/2}\left(\frac{\Delta}{\sqrt{2\pi}}+\frac{1+\Delta^2}{2}\right)\right)+\frac{2\Delta^2}{(2^{B}-2)^2}\nonumber \\
%&=\frac{1}{2}\left(e^{-2B_1 \ln2}+e^{-2B_2 \ln2}\right)+\frac{2-e^{-2B_1 \ln2}}{(2^{B_{1}}-2)^2}+\frac{2-e^{-2B_2 \ln2}}{(2^{B_{2}}-2)^2}\nonumber \\
&\overset{(a)}{\leq} K_{1}e^{-2B \ln2},
\end{align}
\begin{align} \label{ineq:distch_1}
D_{e,2}&< 2 \left(4\Delta^2 \Pr(|u_j|<\Delta)+\int_{\Delta}^{\infty}(u_j+\Delta)^2 f(u_j)du_j +\int_{-\infty}^{-\Delta}(u_j-\Delta)^2 f(u_j)du_j\right)\nonumber \\
%4\Delta_{1}^2 \Pr(|u_1|<\Delta_1)+ 4\Delta_{2}^2 \Pr(|u_2|<\Delta_2)+2\int_{\Delta_1}^{\infty}(u_1+2\Delta_1)^2 f(u_1)du_1 +2\int_{\Delta_2}^{\infty}(u_2+2\Delta_2)^2 f(u_2)du_2 
&\leq 4\left(2\Delta^2(1-e^{-\Delta^2/2})+e^{-\Delta^2/2}(\Delta\sqrt{2/\pi}+1)+\Delta^2e^{-\Delta^2/2}+2\Delta\left(\frac{1}{\sqrt{2\pi}}+\frac{1-\Delta}{2}e^{-\Delta^2/2}\right)\right) \nonumber \\
&=(32B\ln2+4\sqrt{2B\ln2 / \pi}) +4 e^{-2B\ln2}(1-4B\ln2+2\sqrt{2B\ln2 / \pi})
\end{align}
respectively. Note that in step(a) of (\ref{eq:distquant}) the value of $\Delta$ is substituted and to emphasize the exponential term the rest of the factors are given by the coefficient $K_{1}$ which represents $O(B)$. In the same way, for the distortion caused by channel when both sources are in error regardless of being compatible or incompatible, above bound on $D_{e,2}$ is obtained.
The reconstruction error expressions when only one source is in error are derived for compatible and incompatible pairs, respectively.
\begin{align} \label{ineq:distch}
D_{e,c,1}&<\sum_{n=1}^{2^{B_{j}}}\int_{I_{j,n}}(u_{j}-\hat{u}_{j}(n))^2 f(u_{j})du_{j}+|2\theta^2 \sqrt{1-\rho^2}|^2\nonumber \\
&=\int_{\Delta}^{\infty}(u_j-\Delta)^2 f(u_j)du_j+\int_{-\infty}^{-\Delta}(u_j+\Delta)^2 f(u_j)du_j+\sum_{n=2}^{2^{B}-1}\int_{I_{j,n}}(u_{j}-\hat{u}_{j}(n))^2 f(u_{j})du_{j}+4\theta^2(1-\rho^2) \nonumber \\
&\leq 2 e^{-\Delta^2/2}\left(\frac{\Delta}{\sqrt{2\pi}}+\frac{1+\Delta^2}{2}\right)+\frac{\Delta^2}{(2^{B}-2)^2}+4\theta^2(1-\rho^2)\nonumber \\
&\overset{(b)}{\leq} K_{1}e^{-2 B \ln2}/2+4\theta^2(1-\rho^2),
\end{align}
\begin{align} \label{ineq:distch_ic}
D_{e,ic,1}&<\sum_{n=1}^{2^{B}}\int_{I_{j,n}}(u_{j}-\hat{u}_{j}(n))^2 f(u_{j})du_{j}+\int_{u'_2=\theta}^{\infty}\left(\theta \sqrt{1-\rho^2}+\sqrt{1-\rho^2}u'_2\right)^2 f(u'_2||U'_2|>\theta \sqrt{1-\rho^2})d_{u'_2}\nonumber \\
&=\int_{\Delta}^{\infty}(u_j-\Delta)^2 f(u_j)du_j+\int_{-\infty}^{-\Delta}(u_j+\Delta)^2 f(u_j)du_j \nonumber \\
&+\sum_{n=2}^{2^{B}-1}\int_{I_{j,n}}(u_{j}-\hat{u}_{j}(n))^2 f(u_{j})du_{j}+3\theta^2(1-\rho^2)+(1-\rho^2) \nonumber \\
&\leq 2e^{-\Delta^2/2}\left(\frac{\Delta}{\sqrt{2\pi}}+\frac{1+\Delta^2}{2}\right)+\frac{\Delta^2}{(2^{B}-2)^2}+3\theta^2(1-\rho^2)+(1-\rho^2)\nonumber \\
&\overset{(c)}{\leq} K_{1}e^{-2B\ln2}/2+3\theta^2(1-\rho^2)+(1-\rho^2).
\end{align} Given the symmetry of the normal distribution $D_{e,c,1}$ and $D_{e,ic,1}$ are derived and given in a general form for both sources where $j=1,2$. 
To simplify the calculations, the quantization levels and the number of quantization bins are assumed also to be equal to each other. Thus, the quantization distortion (\ref{eq:distquant}) can be bounded by $D_q \leq K_{1}e^{-2B\ln2}$. Accordingly, both $D_{e,c,1}$ and $D_{e,ic,1}$ compose the quantization distortion on one source since they represent one correctly and one incorrectly decoded, this is why in steps (b) and (c) the upper bound on $D_q$ derived in (\ref{eq:distquant}) is used for the source which is decoded correctly.